\documentclass[a4paper,11pt]{article}
\pdfoutput=1

\usepackage{jheppub}
\usepackage{tikz}
\usepackage{verbatim}
\usetikzlibrary{shapes}
\usetikzlibrary{plotmarks}
\usepackage{tikz-3dplot}
\usepackage{footmisc}
\usepackage{ytableau}

\definecolor{light}{rgb}{0,1,1}

\title{Exact Results for SYM on $Y^{p,q}$ and $S^2\times S^2$ with Conical Singularities}
\author[1]{Lorenzo Ruggeri}
\affiliation[1]{Yau Mathematical Sciences Center, Tsinghua University, Beijing, 100084, China}
\emailAdd{ruggeri@mail.tsinghua.edu.cn}
\preprint{}

\abstract{Starting from a theory on $S^3\times S^3$ and dimensionally reducing, we compute the full partition function, including flux and instanton contributions, for an $\mathcal{N}=1$ theory of vector multiplets and hypermultiplets on five-dimensional toric Sasakian manifolds $Y^{p,q}$. Dimensionally reducing, we obtain the partition function for Pestun-like theories on a class of manifolds whose topology is $S^2\times S^2$. Generalizing the procedure starting from branched covers of $S^3\times S^3$, we reduce to a theory on $Y^{p,q}$ with codimension two twist defects. Exploiting a proposed equivalence with partition functions on spaces with orbifold singularities, our results provide the partition function of an $\mathcal{N}=2$ theory on the product of two spindles.}

\begin{document}

\maketitle
\flushbottom

\section{Introduction}
The study of supersymmetric quantum field theories (SQFTs) on non-trivial geometric backgrounds, pioneered in \cite{Pestun:2007rz,Festuccia:2011ws}, has provided profound insights into both physical phenomena and mathematical structures. Among these, five-dimensional toric Sasakian geometries \cite{Fulton:1993,Boyer:2000,Boyer:2008era} have emerged as a particularly rich setting due to their applications in the AdS/CFT correspondence \cite{Klebanov:1998hh,Morrison:1998cs,Martelli:2004wu}, their connection to refined topological strings \cite{Lockhart:2012vp} and as a tool to compute partition function of 4d theories via dimensional reduction \cite{Festuccia:2019akm,Lundin:2021zeb,Lundin:2023tzw}. In \cite{Kallen:2012cs,Imamura:2012efi,Kim:2012qf}, employing supersymmetric localization \cite{Pestun:2007rz,Pestun:2016zxk}, the partition function of an $\mathcal{N}=1$ SQFT has been computed on the simplest instance of toric Sasakian manifold, the five-sphere $S^5$. Later, these results have been extended in \cite{Qiu:2013pta,Qiu:2013aga} to manifolds which are topologically $S^2\times S^3$, namely $Y^{p,q}$ \cite{Martelli:2004wu} and $L^{a,b,c}$ \cite{Cvetic:2005ft}. Moreover, in \cite{Pan:2013uoa,Imamura:2014ima,Alday:2015lta,Pini:2015xha} five-dimensional rigid supersymmetric backgrounds have been classified by coupling to different supergravity multiplets and taking the rigid limit. Eventually, the partition function on a generic toric Sasakian manifold has been obtained in \cite{Qiu:2014oqa,Qiu:2016dyj}.

In these results, the gauge configurations considered are either trivial connections or contact instantons. However, whenever a manifold has non-trivial two-cycles, it is expected that gauge configurations with flux are allowed and the partition function is a sum over topological sector labelled by their first Chern class. For instance, such gauge configurations are expected on $Y^{p,q}$ and $L^{a,b,c}$, as $H_2(S^2\times S^3,\mathbb{Z})\simeq\mathbb{Z}$. One of the main results of this work is to compute their contributions to the partition function. Hence, we are able to provide the full (integrand of the) partition function for an $\mathcal{N}=1$ vector multiplet coupled to a hypermultiplet in a representation $R$ of the gauge group on both $Y^{p,q}$ and $L^{a,b,c}$.

Focusing first on the trivial contact instanton sector, we compute the one-loop determinant around flux configurations. The idea is to uplift the cohomological complex obtained via localization on $Y^{p,q}$ (or $L^{a,b,c}$) to $S^3\times S^3$ \cite{Qiu:2013pta} and compute the equivariant index of the resulting complex in 6d. The index, which provides the perturbative partition function of an $\mathcal{N}=(1,0)$ theory on $S^3\times S^3$, can be decomposed in contributions from different representations under a freely acting $S^1_Y$. Keeping only the trivial representation, the authors of \cite{Qiu:2013pta} computed the perturbative partition function on\footnote{In spirit, this dimensional reduction is similar to that appearing in \cite{Imamura:2012efi} where the partition function of an $\mathcal{N}=1$ theory on $S^5$ is obtained from the index of a 6d $\mathcal{N}=(1,0)$ theory.} $(S^3\times S^3)/S^1_Y\simeq Y^{p,q}$. It has been recently shown in\footnote{See \cite{Benini:2012ui} for an earlier result, where the large $h$ limit of the $\mathcal{N}=2$ partition function on lens spaces $L(h,-1)$ has been shown to reproduce the $\mathcal{N}=(2,2)$ one-loop determinant around fluxes on $S^2$.} \cite{Lundin:2021zeb,Lundin:2023tzw} that, when dimensionally reducing along a non-trivially fibered $S^1$, modes with non-zero charge for rotations under the $S^1$ contribute to different flux sectors on the base manifold. Hence, to compute the one-loop determinant at a given flux sector on $Y^{p,q}$ we simply have to restrict the 6d equivariant index to contributions with a given representation under $S^1_Y$. Rather than shrinking the radius of $S^1_Y$, we perform the dimensional reduction by introducing a finite quotient $(S^3\times S^3)/\mathbb{Z}_h$ and then taking the large $h$ limit. Let us stress that the 6d $\mathcal{N}=(1,0)$ theory is an auxiliary tool which serves as a setup where, due to the absence of non-trivial two-cycles, computing the one-loop determinant, or the equivariant index, is simpler.

The manifold $Y^{p,q}$ is a non-trivial principal bundle over a four-manifold $B$ whose topology is that of the product of two spheres and $H_2(B,\mathbb{Z})\simeq\mathbb{Z}^2$. The non-trivial fibration has been exploited in \cite{Lundin:2023tzw} to compute, for an $\mathcal{N}=2$ vector multiplet, the one-loop determinant on $B$ around a single flux starting from the perturbative partition function on $Y^{p,q}$ \cite{Qiu:2013pta}. Since we have computed the one-loop determinant around fluxes in 5d, the dimensional reduction produces the one-loop determinant around both fluxes on $B$. The theory we find is a generalization of Pestun's theory on $S^4$ \cite{Pestun:2007rz} to four manifolds with a Killing vector for a $T^2$-action and isolated fixed points \cite{Festuccia:2018rew,Festuccia:2019akm}. In particular, unlike for equivariant Donaldson-Witten theories \cite{Witten:1988ze}, the field strength localizes to instantons and anti-instantons at different fixed points.

To compute the dependence on fluxes of the instanton part, we first show how in 5d (4d) the one-loop determinant around fluxes can be factorized in contributions from the fixed fibers (fixed points) under the torus action of the manifold \cite{Pasquetti:2011fj,Beem:2012mb}, with fluxes entering as a shift of the Coulomb branch parameter. This behaviour has been shown to hold in \cite{Lundin:2021zeb,Lundin:2023tzw} for a single flux on a large class of quasi-toric four-manifolds. It is conjectured that the same shift encodes also the dependence of the instanton part \cite{Nekrasov:2003vi,Festuccia:2018rew,Festuccia:2019akm}. Under this assumption, we can write down the full partition function\footnote{Note that full partition functions for equivariant Donaldson-Witten theories on complex toric surfaces have been studied also in \cite{Bershtein:2015xfa,Bershtein:2016mxz,Bonelli:2020xps}. However, in that case fluxes are equivariant fluxes further constrained by stability conditions. In our work all expressions are in terms of physical fluxes.}, including contact instantons on $Y^{p,q}$ and $L^{a,b,c}$, and instantons on $B$.

Another class of spaces that has drawn a lot of attention recently, following the finding of black hole solutions with near horizon geometries containing orbifold singularities \cite{Ferrero:2020laf,Ferrero:2020twa,Ferrero:2021etw}, is that of weighted projective spaces $\mathbb{CP}^r_{\boldsymbol{p}}$, also known as spindles in 2d. Recently \cite{Colombo:2024mts}, the partition function of an holographic $\mathcal{N}=2$ theory on $\mathbb{CP}^1_{\boldsymbol{p}}\times S^1$ \cite{Inglese:2023wky,Inglese:2023tyc} has been shown to reproduce the entropy of an AdS$_4$ black hole with spindle horizon, paving the way for studying the AdS/CFT correspondence in a setup with singularities.

These recent developments demand for a more detailed understanding of SQFTs on spaces with orbifold singularities. In \cite{Mauch:2024uyt}\footnote{See also \cite{Martelli:2023oqk} for similar results concerning the equivariant index on $\mathbb{CP}^2_{\boldsymbol{p}}$.}, the full partition function on $\mathbb{CP}^r_{\boldsymbol{p}}$ has been shown to be reproduced by a certain theory on $\mathbb{CP}^2$ obtained dimensionally reducing from a branched cover of $S^{2r-1}$. Theories on branched covers can be equivalently described as theories on smooth manifolds with the insertion of codimension two twist defects \cite{Dixon:1986qv}. The idea behind the equivalence is that the complicated geometry of the branched covers can be traded for fields living on single sheet but with a branching structure in the target space \cite{Cardy:2007mb,Calabrese:2009qy}. The twist defects exactly implement such conditions in the target space. Hence, the theory obtained in \cite{Mauch:2024uyt} on $\mathbb{CP}^r$, and reproducing the partition functions on weighted projective spaces, has insertions of twist defects at the fixed points of the manifold. 

This observation lead to a proposal in \cite{Mauch:2025irx}, for a vector multiplet and a hypermultiplet in the adjoint representation, that theories with twist defects reproduce partition functions on more generic orbifolds\footnote{In \cite{Mauch:2025irx}, it is considered also the insertion of Gukov-Witten defects which are needed to embed the theories with twist defects into a parent theory with larger gauge group. Such embedding, while not affecting the partition functions, is performed to make contact with earlier works studying Gukov-Witten defects and orbifolds \cite{Kanno:2011fw}. In this work we will not study their insertion, however it can always be performed along the lines of \cite{Mauch:2025irx}. It remains an open problem to extend this construction to matter in a generic representation of the gauge group.}. Therefore, we start from the equivariant index, or perturbative partition function, on the branched cover of $S^3\times S^3$, and repeating the previous steps, we compute the full partition function for theories with twist defects on $Y^{p,q}$, $L^{a,b,c}$ and $B$. According to the relation proposed in \cite{Mauch:2024uyt,Mauch:2025irx}, the expressions we find reproduce the partition functions on the corresponding spaces with orbifold singularities. In particular, in 4d we propose the partition function for an $\mathcal{N}=2$ theory on a space whose topology is that of the product of two spindles.

The outlook is as follows. In \autoref{sec.2} we introduce some basic facts about toric Sasakian geometry, explaining in some details the dimensional reductions from $S^3\times S^3$, and its branched cover, down to $Y^{p,q}$ and $B$. We further discuss the corresponding spaces with deficit angles. Then, in \autoref{sec.3}, we present the perturbative partition function on the five-dimensional spaces, also in the setup with twist defects. In \autoref{sec.4} we perform the dimensional reduction, first from 6d to 5d and then from 5d to 4d, which enables us to compute the one-loop determinant around fluxes on these spaces. The factorisation properties of the partition functions are studied in \autoref{sec.5} where, employing the shifts derived from the one-loop part around fluxes, we can write the full partition function, including instanton contributions. We summarise our results in \autoref{sec.6} and comment on possible future directions.

\section{Geometric Setup}\label{sec.2}
In this section, we briefly recall the geometric properties of two families of toric Sasakian manifolds, namely $Y^{p,q}$ \cite{Martelli:2004wu} and $L^{a,b,c}$ \cite{Cvetic:2005ft}, whose topology is $S^2\times S^3$. We refer to \cite{Fulton:1993,Boyer:2000,Boyer:2008era,Sparks:2010sn} for a detailed account of Sasakian and toric geometry. Here, we focus on showing how both $Y^{p,q}$ and $L^{a,b,c}$ arise via free $S^1$-quotients of $S^3\times S^3$. We will also introduce $B=Y^{p,q}/S^1$, a class of quasi-toric manifolds whose topology is $S^2\times S^2$. We will exploit these fibrations to compute the one-loop determinant at all flux sectors on these manifolds in \autoref{sec.4}.

In the following sections, we will also study partition functions on spaces with orbifold singularities and underlying manifolds $Y^{p,q}$, $L^{a,b,c}$ and $S^2\times S^2$. To obtain these we will consider dimensional reduction from a branched cover of $S^3\times S^3$. Hence, we conclude this section by describing the geometries associated to these branched covers and orbifolds.

\subsection{Toric Sasakian Manifolds}
Sasakian manifolds $(M,g)$ are defined as Riemannian manifolds whose metric cone $(C(M),\bar{g})$ is K\"ahler
\begin{equation}\label{eq.metric}
    C(M)=M\times\mathbb{R}_{> 0},\qquad \bar{g}=dr^2+r^2g.
\end{equation}
The existence of an effective, holomorphic and Hamiltonian torus action on the K\"ahler cone leads to the definition of toric Sasakian manifolds. Moreover, $C(M)$ can be obtained as a K\"ahler quotient
\begin{equation}\label{eq.Kquotient}
    C(M)=\mathbb{C}^m// U(1)^{m-3},
\end{equation}
where $U(1)^{m-3}$ acts with charges
\begin{equation}
    \vec{Q}^M_a=[Q^1_a,\dots,Q^m_a],\qquad a=1,\dots,m-3, 
\end{equation}
on $\mathbb{C}^m$. If the charges satisfy the $m-3$ conditions $Q^1_a+\dots+Q^m_a=0$, the cone is Calabi-Yau and $M$ is said to be Sasaki-Einstein.

Restricting to five-dimensional simply connected toric Sasakian manifolds, the simplest example, with $m=3$, is $S^5$ whose metric cone is $\mathbb{C}^3/\{0\}$. Moving to $m=4$, there are two classes of manifolds, $Y^{p,q}$ and $L^{a,b,c}$, whose metric cones are defined by the charges of the $U(1)$ in \eqref{eq.Kquotient}
\begin{equation}\begin{split}\label{eq.chargesYL}
    \mathbb{C}^m//S^1_Y=C(Y^{p,q}):\qquad& \vec{Q}^Y=[p+q,p-q,-p,-p],\quad p>q>0,\\
    \mathbb{C}^m//S^1_L=C(L^{a,b,c}):\qquad& \vec{Q}^L=[a,b,-c,-a-b+c],\;\quad a,b,a+b-c>0,
\end{split}\end{equation}
where $\gcd(p,q)=\gcd(a,b)=\gcd(a,c)=\gcd(b,c)=1$. Each component of these four-vectors rotates the phase of each $\mathbb{C}$ factor. As the sum of the charges in \eqref{eq.chargesYL} vanishes in both cases, these two families of manifolds are Sasaki-Einstein. 

Let us explicitly construct the class of $Y^{p,q}$ manifolds starting from $\mathbb{C}^4$, following the presentation in \cite{Qiu:2013pta,Qiu:2016dyj}. The K\"ahler quotient consists in 
\begin{equation}\label{eq.mu-1/S1}
    C(Y^{p,q})=\mu_{S^1_Y}^{-1}(0)/S^1_Y,
\end{equation}
where the moment map for $S^1_Y$ is given by 
\begin{equation}\label{eq.momentY}
    \mu_{S^1_Y}=(p+q)|z_1|^2+(p-q)|z_2|^2-p|z_3|^2-p|z_4|^2.
\end{equation}
To further obtain $Y^{p,q}$ from \eqref{eq.mu-1/S1} we need to fix the scaling freedom in \eqref{eq.metric}, which we achieve by restricting to a hypersurface
\begin{equation}
    H_\omega=\{z_i\in\mathbb{C}\,|\,\sum_{i=1}^4\omega_i|z_i|^2=1\},
\end{equation}
where $\omega_i\in\mathbb{R}$. We choose all $\omega_i>0$, so that the intersection $\mu_{S^1_Y}^{-1}(0)\cap H_\omega$ is topologically $S^3\times S^3$. Moreover, as both the loci $z_1=z_2=0$ and $z_3=z_4=0$ are not part of the intersection, the $S^1_Y$-action acts freely on $\mu_{S^1_Y}^{-1}(0)\cap H_\omega$. Hence, we find
\begin{equation}
    Y^{p,q}=(\mu_{S^1_Y}^{-1}(0)\cap H_\omega)/S^1_Y.
\end{equation}
A possible choice for the hypersurface is the following
\begin{equation}
    (p+q)|z_1|^2+(p-q)|z_2|^2+p|z_3|^2+p|z_4|^2=1,
\end{equation}
so that $S^3\times S^3\simeq \mu_{S^1_Y}^{-1}(0)\cap H_\omega$ is given by
\begin{equation}\label{eq.S3S3}
    S^3\times S^3\simeq\{(z_1,z_2,z_3,z_4)\in\mathbb{C}^4\,|\,(p+q)|z_1|^2+(p-q)|z_2|^2=\frac{1}{2},\; p|z_3|^2+p|z_4|^2=\frac{1}{2}\}.
\end{equation}
Each $S^3$ can be seen as a $T^2$-fibration over an interval. Hence, $S^3\times S^3$ can be seen as a $T^4$-fibration over a polytope with four facets. The choice in \eqref{eq.S3S3} has, as toric diagram, exactly the one appearing on the left hand side in \autoref{fig.toricY}.

The same construction can be applied to $L^{a,b,c}$ manifolds, simply by using the charges $\vec{Q}^L$ in \eqref{eq.chargesYL} and modifying the hypersurface $H_\omega$ accordingly \cite{Qiu:2013pta}. Hence, also the product of three-spheres in \eqref{eq.S3S3} will have to be modified. The precise definition of these objects will not be necessary to compute the partition function in the next section. However, the toric diagram of the corresponding $S^3\times S^3$ is, by construction, the one appearing on the right hand side in \autoref{fig.toricY}.

Exploiting the toric structure, both $Y^{p,q}$ and $L^{a,b,c}$ can be seen as a Lagrangian $T^3$-fibration over their corresponding moment polytopes \cite{Lerman:2001zua}, shown in \autoref{fig.toricY}.
\begin{figure}
\centering
\begin{tikzpicture}[scale=1]
    \draw[line width=1.2pt] (2,0)--(3,0)--(2,4)--(1,1)--cycle;
    \draw[step=1.0,gray!60] (-.2,-.2) grid (4.2,4.2);
    \node at (2,0) [below]{$(0,0)$};
    \node at (3,0) [below]{$(1,0)$};
    \node at (2,4) [left]{$(0,4)$};
    \node at (1,1) [left]{$(-1,1)$};
    \node at (1.5,0.5) [left]{$C_1$};
    \node at (2.5,0) [above]{$C_2$};
    \node at (2.5,2) [right]{$C_3$};
    \node at (1.3,2) [left]{$C_4$};  
    \begin{scope}[shift={(7,0)}]
    \draw[line width=1.2pt] (2,0)--(3,0)--(3,1)--(1,4)--cycle;
    \draw[step=1.0,gray!60] (-.2,-.2) grid (4.2,4.2);
    \node at (2,0) [below]{$(0,0)$};
    \node at (3,0) [below]{$(1,0)$};
    \node at (3,1) [right]{$(1,1)$};
    \node at (1,4) [left]{$(-1,4)$};
    \node at (1,2) {$C_1$};
    \node at (2.5,0) [above]{$C_2$};
    \node at (3,0.5) [right]{$C_3$};  
    \node at (2.4,2) [right]{$C_4$};
    \end{scope}
\end{tikzpicture}
\caption{Toric diagrams for $Y^{p,q}$ and $L^{a,b,c}$. Left side: $Y^{4.3}$. Right side: $L^{1,5,2}$. The facets $C_i$ label the codimension two submanifolds kept fixed by an $S^1$ in the $T^3$ Cartan isometry. The four fixed fibers of each manifold are at the intersection of two facets.}
\label{fig.toricY}
\end{figure}
In both cases, each of the four facets $C_i$ represents a codimension two submanifold where a circle subgroup of the $T^3$ Cartan isometry degenerates. In particular, for $Y^{p,q}$, these are given by lens spaces
\begin{equation}\label{eq.lensfacets}
    C_1\cong C_2\cong S^3/\mathbb{Z}_{p},\qquad C_3\cong S^3/\mathbb{Z}_{p+q},\qquad C_4\cong S^3/\mathbb{Z}_{p-q}.
\end{equation}
There are also four codimension four submanifolds which are fixed under a $T^2\subset T^3$ action. These are given by the intersection of two facets $C_i\cap C_{i+1}$ where, locally, the manifold looks like $\mathbb{C}^2\times S^1$.

Whenever a toric Sasakian manifolds $M$ admits a globally free $S^1$-action, the quotient leads to a quasi-toric four-manifold $M/S^1$  \cite{Festuccia:2019akm,Lundin:2021zeb,Lundin:2023tzw}. As $L^{a,b,c}$ does not admit any free $S^1$-action, we focus on $Y^{p,q}$. This manifold is a principal $S^1$-bundle over the product of an axially squashed and a round sphere $B=\mathbb{CP}^1\times\mathbb{CP}^1$, such that the first Chern number on the standard two-cycles are $p$ and $q$.

To explicitly write the fiber of $Y^{p,q}$, we perform a change of basis which better suits this setup with a reduced torus action. So far, we wrote four-vectors in terms of a basis $[e_1,e_2,e_3,e_4]$, where each $e_i$ is a vector field generating rotations of the phase of each $\mathbb{C}$ factor. Now, we employ a new basis $[\tilde{e}_1,\tilde{e}_2,\tilde{e}_3,e_u]$ where the first three components generate the $T^3$-action in $Y^{p,q}$ and $e_u$ generates the free $S^1_Y$-action \eqref{eq.chargesYL} in $S^3\times S^3$. Thus, when studying the five-dimensional setup, we can simply forget about the last component and write vectors in terms of $[\tilde{e}_1,\tilde{e}_2,\tilde{e}_3]$. This change of basis, which appears in section 3.2 in \cite{Qiu:2014oqa}, will be presented explicitly in the next section when we will rewrite the perturbative partition function on $Y^{p,q}$.

As $Y^{p,q}$ manifolds are at most quasi-regular Sasakian manifolds \cite{Sparks:2010sn}, the Reeb vector does not generate a globally free $S^1$-action. However, there is a globally free $S^1$-action which is not parallel to the Reeb vector at all fixed fibers\footnote{The exception is $T^{1,1}$, whose metric cone is the conifold, and that corresponds to $p=1$ and $q=0$. This manifold is regular and in this case one can also consider an $S^1$-action parallel to the Reeb at all fixed fibers.}, given by \cite{Festuccia:2016gul,Lundin:2023tzw}
\begin{equation}\label{eq.freeS1Y}
    \vec{\X}^\text{ex}=[1,0,0].
\end{equation}
The quotient $B=Y^{p,q}/S^1$ can be seen as a $T^2$-fibration over the polytope on the left-hand side in \autoref{fig.toricY}. In this case, the facets label different $\mathbb{CP}^1$'s and their four intersections label the fixed points of the manifold.

\subsection{Branched Covers and Orbifolds}
We now introduce branched covers of $S^3\times S^3$ and orbifolds in 5d and 4d whose underlying manifolds are $Y^{p,q}$, $L^{a,b,c}$ and $B$. We denote $\boldsymbol{p}=(p_1,p_2,p_3,p_4)$ and $|\boldsymbol{p}|=p_1p_2p_3p_4$.

\paragraph{Branched Covers.}
The starting point is a branched cover of $S^3\times S^3$, which can be embedded in a branched cover of $\mathbb{C}^4$, with coordinates $\tilde{z}_i=\rho_ie^{\ii\tilde{\theta}_i}$ and angles $\tilde{\theta}_i\in[0,2\pi p_i)$, as follows
\begin{equation}\label{eq.S3S3alpha}
    \widehat{S}^3_{\boldsymbol{p_A}}\times \widehat{S}^3_{\boldsymbol{p_B}}\simeq\{(\tilde{z}_1,\tilde{z}_2,\tilde{z}_3,\tilde{z}_4)\in\widehat{\mathbb{C}}_{\boldsymbol{p}}^4\,|\,(p+q)|\tilde{z}_1|^2+(p-q)|\tilde{z}_2|^2=\frac{1}{2},\; p|\tilde{z}_3|^2+p|\tilde{z}_4|^2=\frac{1}{2}\}.
\end{equation}
where we labelled $\boldsymbol{p}_A=(p_1,p_2)$ and $\boldsymbol{p}_B=(p_3,p_4)$.

Also in this case, each individual branched cover of $S^3$ can be seen as a $T^2$-fibration over an interval. Hence, $\widehat{S}^3_{\boldsymbol{p_A}}\times \widehat{S}^3_{\boldsymbol{p_B}}$ can be seen as a $T^4$-fibration over a polytope with four facets, shown on the left hand side of \autoref{fig.toricY.alpha}.
\begin{figure}
\centering
\begin{tikzpicture}[scale=1]
    \draw[line width=1.2pt] (2,0)--(3,0)--(2,4)--(1,1)--cycle;
    \draw[step=1.0,gray!60] (-.2,-.2) grid (4.2,4.2);
    \node at (2,0) [below]{$p_1p_3$};
    \node at (3,0) [below]{$p_3p_2$};
    \node at (2,4) [left]{$p_2p_4$};
    \node at (1,1) [left]{$p_4p_1$};
    \node at (1.5,0.5) [left]{$p_1$};
    \node at (2.5,0) [above]{$p_3$};
    \node at (2.5,2) [right]{$p_2$};
    \node at (1.3,2) [left]{$p_4$};  
    \begin{scope}[shift={(7,0)}]
    \draw[line width=1.2pt] (2,0)--(3,0)--(3,1)--(1,4)--cycle;
    \draw[step=1.0,gray!60] (-.2,-.2) grid (4.2,4.2);
    \node at (2,0) [below]{$p_1p_3$};
    \node at (3,0) [below]{$p_3p_2$};
    \node at (3,1) [right]{$p_2p_4$};
    \node at (1,4) [left]{$p_4p_1$};
    \node at (1,2) {$p_1$};
    \node at (2.5,0) [above]{$p_3$};
    \node at (3,0.5) [right]{$p_2$};  
    \node at (2.4,2) [right]{$p_4$};
    \end{scope}
\end{tikzpicture}
\caption{Left side: toric diagram and singularity structure of $\widehat{S}^3_{\boldsymbol{p_A}}\times \widehat{S}^3_{\boldsymbol{p_B}}$ in \eqref{eq.S3S3alpha}. Right side: toric diagram and singularity structure of a different $\widehat{S}^3_{\boldsymbol{p_A}}\times \widehat{S}^3_{\boldsymbol{p_B}}$ such that its quotient gives $\widehat{L}^{a,b,c}_{\boldsymbol{p}}$.}
\label{fig.toricY.alpha}
\end{figure}
In the interior of the polytope, where $\tilde{z}_i\neq 0,\;\forall\,i$, this space is a $|\boldsymbol{p}|$-fold cover of $S^3\times S^3$. The facets of the square, which are codimension four submanifolds, contain conical singularities. Each facets is characterized by a different $\tilde{z}_i$ vanishing. Here, $p_i$ sheets collapse and the space is a $|\boldsymbol{p}|/p_i$-fold cover of $S^3\times S^3$ with branching index $p_i$. At the codimension four locus, where two facets meet, a pair of $\tilde{z}_i$ vanish. Considering for instance $\tilde{z}_1=\tilde{z}_3=0$, here the space is a $|\boldsymbol{p}|/p_1p_3$-fold cover of $S^3\times S^3$ with branching index $p_1p_{3}$. Note that as $z_1,z_2$ or $z_3,z_4$ cannot vanish simultaneously, branching indices $p_1p_2$ or $p_3p_4$ do not appear.

We consider quotients by locally free $S^1_{\widehat{Y}}$ and $S^1_{\widehat{L}}$ actions with charges 
\begin{equation}\begin{split}\label{eq.chargesYL.alpha}
    \vec{Q}^{\widehat{Y}}=&\left[\frac{p+q}{p_1},\frac{p-q}{p_2},-\frac{p}{p_3},-\frac{p}{p_4}\right],\\
    \vec{Q}^{\widehat{L}}=&\left[\frac{a}{p_1},\frac{b}{p_2},-\frac{c}{p_3},-\frac{a+b-c}{p_4}\right].
\end{split}\end{equation}
The action is scaled with respect to \eqref{eq.chargesYL} to take into account the larger periodicities of the angles $\tilde{\theta}_i$. Along the lines of \cite{Mauch:2024uyt}, where it has been shown that the $S^1$-quotients of $\widehat{S}^{2r-1}_{\boldsymbol{p}}$ give rise to smooth complex projective spaces, one can show that
\begin{equation}
    (\widehat{S}^3_{\boldsymbol{p_A}}\times \widehat{S}^3_{\boldsymbol{p_B}})/S^1_{\widehat{Y}}=\widehat{Y}^{p,q}_{\boldsymbol{p}}.
\end{equation}
The base $\widehat{Y}^{p,q}_{\boldsymbol{p}}$ is simply $Y^{p,q}$. Despite this fact, we stick with the notation $\widehat{Y}^{p,q}_{\boldsymbol{p}}$ for the base space, signifying that the resulting partition function on $\widehat{Y}^{p,q}_{\boldsymbol{p}}$ is not the one on $Y^{p,q}$. As we will discuss shortly, the theory on $\widehat{Y}^{p,q}_{\boldsymbol{p}}$ includes twist defects \cite{Dixon:1986qv}.

As for the smooth case considered above, this procedure can be applied also to obtain $\widehat{L}^{a,b,c}_{\boldsymbol{p}}$, simply employing the charge $\vec{Q}^{\widehat{L}}$ in \eqref{eq.chargesYL.alpha}. The polytope and the singularity structure of the corresponding product of branched covers of $S^3$ is shown on the right hand side of \autoref{fig.toricY.alpha}.

We will also consider the quotient $\widehat{Y}^{p,q}_{\boldsymbol{p}}/S^1=\widehat{B}_{\boldsymbol{p}}$, where the $S^1$-action is that in \eqref{eq.freeS1Y}. Let us stress again that, despite the different notation, $\widehat{B}_{\boldsymbol{p}}$ is a smooth manifold whose topology is that of $S^2\times S^2$.

\paragraph{Orbifolds.}
In \cite{Mauch:2024uyt}, it is shown that the partition functions of a theory on $\mathbb{CP}^r$ obtained dimensionally reducing from branched covers of $S^{2r-1}$, match those computed on weighted projective spaces $\mathbb{CP}^r_{\boldsymbol{p}}$. A required condition is that the singularity structure of the branched covers of the sphere match that of the weighted projective space. 

Hence, we introduce orbifolds $Y^{p,q}_{\boldsymbol{p}}$ and $L^{a,b,c}_{\boldsymbol{p}}$ which are a $T^3$-fibration over the toric diagrams in \autoref{fig.toricY.alpha}. As these polytopes match those in \autoref{fig.toricY} for $Y^{p,q}$ and $L^{a,b,c}$, in the interior these orbifolds are identical to the corresponding manifolds. However, in the interior of each facet $C_i$ there is an orbifold singularity and the space locally looks like $(\mathbb{C}/\mathbb{Z}_{p_i})\times\mathbb{C}\times S^1$. At the intersection of two facets, the space is locally $(\mathbb{C}/\mathbb{Z}_{p_i})\times(\mathbb{C}/\mathbb{Z}_{p_{i+1}})\times S^1$. The facets are given by lens spaces with orbifold singularities
\begin{equation}\begin{split}
    &C_1=S^3_{(p_4p_1,p_1p_3)}/\mathbb{Z}_p,\quad\qquad C_2=S^3_{(p_1p_3,p_3p_2)}/\mathbb{Z}_p,\\
    &C_3=S^3_{(p_3p_2,p_2p_4)}/\mathbb{Z}_{p+q},\quad\quad C_4=S^3_{(p_2p_4,p_4p_1)}/\mathbb{Z}_{p-q}.
\end{split}\end{equation}
The singularity structure of these orbifolds is chosen such that it reproduces the branching indices of $\widehat{S}^3_{\boldsymbol{p_A}}\times \widehat{S}^3_{\boldsymbol{p_B}}$ (or of the product of branched covers of three-spheres corresponding to $\widehat{L}^{a,b,c}_{\boldsymbol{p}}$).

Upon dimensional reduction from $\widehat{Y}^{p,q}_{\boldsymbol{p}}$, we will find a theory on $\widehat{B}_{\boldsymbol{p}}$ whose partition function we expect to reproduce that on an orbifold $B_{\boldsymbol{p}}$. The toric diagram and singularity structure of these orbifolds is the one appearing on the left side in \autoref{fig.toricY.alpha} and, exactly as $B$, they are a $T^2$-fibration over the interior of the diagram. At the facets, where an $S^1\subset T^2$ degenerates, there is an orbifold singularity. Similarly at the insertion of two facets.

\section{5d Partition Function}\label{sec.3}

In this section we consider an $\mathcal{N}=1$ vector multiplet and a hypermultiplet in a representation $R$ of the gauge group $G$ on two classes of toric Sasakian manifolds, $Y^{p,q}$ and $L^{a,b,c}$. The perturbative partition function has been computed in \cite{Qiu:2013pta} dimensionally reducing from an $\mathcal{N}=(1,0)$ theory on $S^3\times S^3$. We start by briefly summarizing their derivation, which we then generalize starting instead from branched covers $\widehat{S}^3_{\boldsymbol{p_A}}\times \widehat{S}^3_{\boldsymbol{p_B}}$, to compute the perturbative partition function on $\widehat{Y}^{p,q}_{\boldsymbol{p}}$ and $\widehat{L}^{a,b,c}_{\boldsymbol{p}}$. Via the identification of theories on branched covers and theories on smooth manifolds with the insertion of twist defects \cite{Calabrese:2009qy}, the partition function we are computing on $\widehat{Y}^{p,q}_{\boldsymbol{p}}$ and $\widehat{L}^{a,b,c}_{\boldsymbol{p}}$ correspond to the insertion of twist defects on these manifolds. This can be explicitly shown by translating the theory on $\widehat{S}^3_{\boldsymbol{p_A}}\times \widehat{S}^3_{\boldsymbol{p_B}}$ as one on $S^3\times S^3$ with twist defects and then dimensionally reducing along the $S^1$-action in \eqref{eq.chargesYL}. We will not discuss this equivalence here and refer to \cite{Calabrese:2009qy,Giveon:2015cgs,Nishioka:2016guu,Mauch:2025irx}. In particular \cite{Nishioka:2016guu} explicitly shows the equivalence for branched covers of $S^3$. Their argument can be easily adapted to $\widehat{S}^3_{\boldsymbol{p_A}}\times \widehat{S}^3_{\boldsymbol{p_B}}$.

As just argued, the $\mathcal{N}=1$ theories on $\widehat{Y}^{p,q}_{\boldsymbol{p}}$ and $\widehat{L}^{a,b,c}_{\boldsymbol{p}}$ are found by placing twist defects in the $\mathcal{N}=1$ theories on  $Y^{p,q}$ and $L^{a,b,c}$. Such theories can be studied for a vector multiplet and a hypermultiplet in a generic representation of the gauge group. Thus, in the rest of the paper when presenting the partition functions we will consider hypermultiplets in a generic representation of the gauge group.

A crucial point is that for vector multiplets and hypermultiplets in the adjoint of $G$, following the proposal in \cite{Mauch:2024uyt,Mauch:2025irx}, all the results we compute on $\widehat{Y}^{p,q}_{\boldsymbol{p}}$ and $\widehat{L}^{a,b,c}_{\boldsymbol{p}}$ hold also for the corresponding spaces $Y^{p,q}_{\boldsymbol{p}}$ and $L^{a,b,c}_{\boldsymbol{p}}$ with orbifold singularities introduced in \autoref{sec.2}. As explained in the introduction, the proposal relies on (i) the the results of \cite{Mauch:2024uyt} where it has been shown that these theories reproduce all known partition functions computed on orbifolds \cite{Inglese:2023wky,Inglese:2023tyc,Martelli:2023oqk}, and (ii) earlier results studying orbifolds and codimension two defects \cite{Kanno:2011fw}. It remains an open problem\footnote{As explained in \cite{Mauch:2025irx}, the problem for hypermultiplets in a generic representation is that the effects of Gukov-Witten and twist defects do not seem to be compatible. To extend the equivalence with theories on orbifolds one can either find a way to make the combined effect compatible or argue that Gukov-Witten defects are not needed for the equivalence. We hope to address this problem in future work.} to establish whether this equivalence holds for matter in a generic representation of $G$.

We will work with cohomologically twisted theories throughout this work, see \cite{Kallen:2012cs} for a definition of the cohomological variables and the corresponding complex. Defining cohomological variables for hypermultiplets \cite{Festuccia:2020yff} requires the existence of a spin, or a spin$^c$, structure. The toric Sasakian manifolds we consider, and the 4d quotients, are spin. Hence, one can switch between the physical and cohomologically twisted theories bijectively. Moreover, all branched covers and orbifolds we will consider are topologically a product space of two-spheres and three-spheres with conical singularities associated to either surplus or deficit angles. A spin structure can be defined on the spindle \cite{Ferrero:2021etw}, on orbifolds of $S^3$ \cite{Inglese:2023tyc} and on branched covers of $S^3$ \cite{Nishioka:2013haa}. Therefore, also the product spaces we consider in 4d, 5d and 6d are spin.

\subsection{Localization}
The full partition function on generic toric Sasakian manifold $\widehat{M}_{\boldsymbol{p}}$, possibly with the insertion of twist defects\footnote{The case $\boldsymbol{p}=(1,\dots,1)$ corresponds to a manifold $M$ with no insertion of twist defects.}, is given by
\begin{equation}\label{eq.Zfull.M}
    \mathcal{Z}_{\widehat{M}_{\boldsymbol{p}}}=\sum_{\mathfrak{m}_1,\dots,\mathfrak{m}_{b_2}}\int_{\mathfrak{h}}\dd a\, e^{-S_\text{cl}}\cdot Z_{\widehat{M}_{\boldsymbol{p}}}\cdot Z_{\widehat{M}_{\boldsymbol{p}}}^\text{inst}.
\end{equation}
The sum is over gauge fluxes $\mathfrak{m}_i=\text{diag}(m_{i\,1},\dots,m_{i\,\text{rk }G})\in\mathbb{Z}^{\text{rk }G\times\text{rk }G}$ 
\begin{equation}
    \frac{1}{2\pi}\int_{\Sigma_i}F=\mathfrak{m}_i,\qquad i=1,\dots,b_2,
\end{equation}
where $b_2$ is the second Betti number of $\widehat{M}_{\boldsymbol{p}}$ and $\Sigma_i$ is a basis of $H_2(\widehat{M}_{\boldsymbol{p}},\mathbb{Z})$. The spaces we consider have non trivial second homology group, $H_2(Y^{p,q},\mathbb{Z})\simeq H_2(L^{a,b,c},\mathbb{Z})\simeq \mathbb{Z}$, and thus we expect a sum over a single flux in the partition function\footnote{Earlier results for the partition function on $Y^{p,q}$ and $L^{a,b,c}$ \cite{Qiu:2013pta,Qiu:2013aga,Qiu:2016dyj} only considered the trivial flux sector.}. Moreover, the integral with respect to the Coulomb branch parameter $a$ is over the Cartan subalgebra $\mathfrak{h}$ of the gauge group $G$. 

The classical contribution is the non $\mathcal{Q}$-exact action evaluated on the BPS locus. At the trivial flux sector, this is given by
\begin{equation}\label{eq.classical.M}
    S_\text{cl}|_{(\mathfrak{m}_1,\dots,\mathfrak{m}_{b_2})=(0,\dots,0)}=-\frac{8\pi^3}{g^2_\text{YM}}\varrho\,\text{tr}(a^2),\qquad\varrho=\frac{\text{Vol}_{\widehat{M}_{\boldsymbol{p}}}}{\text{Vol}_{S^5}}.
\end{equation}
The other terms in the integrand are the 1-loop determinant and the contact instanton contributions at each flux sector.

In the absence of twist defects, and restricting to the trivial flux sector, the perturbative partition function is given by \cite{Qiu:2013pta,Qiu:2016dyj}
\begin{equation}\label{eq.1loopdet}
    Z_M|_{(\mathfrak{m}_1,\dots,\mathfrak{m}_{b_2})=(0,\dots,0)}=\frac{\det'_\text{adj}S_3^{\Lambda_M}(\ii a|\vec{\omega})}{\det_{R}S_3^{\Lambda_M}(\ii a+\ii \hat{m}+\frac{1}{2}\bar{\omega}|\vec{\omega})},
\end{equation}
where the numerator and denominator are, respectively, the vector multiplet and hypermultiplet contributions. Moreover, we denoted $\sum_{i=1}^m\omega_i\equiv\bar{\omega}$, and $\hat{m}$, the hypermultiplet mass, enters as a shift or $a$. The generalized triple sine function is defined as follows
\begin{equation}
    S_3^{\Lambda_M}(x|\vec{\omega})=\prod_{\vec{n}\in\Lambda_M}\left(\vec{n}\cdot\vec{\omega}+x\right)\prod_{\vec{n}\in\Lambda_M^\circ}\left(\vec{n}\cdot\vec{\omega}-x\right).
\end{equation}
Here, $\vec{n}=(n_1,\dots,n_m)$ is the vector of charges of the modes under the $T^m$-action of $\mathbb{C}^m$ in \eqref{eq.Kquotient} and $\vec{\omega}=(\omega_1,\dots,\omega_m)$ is the squashing of $M$. Finally, the product over $\vec{n}$ takes values in a lattice
\begin{equation}\label{eq.LambdaM}
    \Lambda_M=\{\vec{n}\in\mathbb{Z}_{\geq 0}^m\,|\,\vec{Q}^M_a\cdot\vec{n}=0,\; a=1,\dots,m-3\}
\end{equation}
and in its interior $\Lambda^\circ_M$. The $m-3$ conditions in the definition of $\Lambda_M$ make sure that the partition function only depends on charges of the modes under the $T^3$-action on $M$. 

Instead of using the $\det'_\text{adj}$-notation, we write the perturbative part of the partition function explicitly as a product over the root set $\Delta$ of the gauge algebra $\mathfrak{g}$
\begin{equation}\label{eq.1looproot}
    Z_M|_{(\mathfrak{m}_1,\dots,\mathfrak{m}_{b_2})=(0,\dots,0)}=\frac{\prod\limits_{\alpha\in\Delta}S_3^{\Lambda_M}(\ii\alpha(a)|\vec{\omega})}{\prod\limits_{\rho\in R}S_3^{\Lambda_M}(\ii a+\ii\hat{m}+\frac{1}{2}\bar{\omega}|\vec{\omega})}.
\end{equation}
Thus, we omit possible factors arising from fermionic zero-modes that would cancel a Vandermonde determinant in the integral over $a$. Momentarily, we will show how to derive \eqref{eq.1looproot} for $Y^{p,q}$ and $L^{a,b,c}$.

As mentioned earlier, the perturbative partition function can be given a cone description by acting with an $SL(m,\mathbb{Z})$ matrix \cite{Qiu:2014oqa}. This is particularly useful in order to reduce the expressions to 4d. For $m=4$, the matrix relevant for $Y^{p,q}$ is given by\footnote{Note that, with respect to \cite{Qiu:2014oqa}, we have flipped the sign of the second and third columns to match the conventions used in \cite{Lundin:2023tzw}.}
\begin{equation}\label{eq.matrix}
   A=\begin{pmatrix}
    0 & 1 & -a & -p-q\\
    0 & 0 & a+2b & -p+q\\
    1 & -1 & -b & p\\
    0 & 0 & -b & p
    \end{pmatrix},\quad (a+b)p+bq=1. 
\end{equation}
By inserting $AA^{-1}$ into $\vec{n}\cdot\vec{\omega}$ we find 
\begin{equation}\label{eq.conelattice}
    \prod_{\vec{n}\in\Lambda_Y}(\vec{n}\cdot\vec{\omega}+x)=\prod_{\vec{n}\in\Lambda_Y}(\vec{n}AA^{-1}\vec{\omega}+x)=\prod_{\vec{m}\in\mathcal{C}_Y}(\vec{m}\cdot\vec{\R}+x),
\end{equation}
where
\begin{equation}\label{eq.m_i}
    m_1=n_3,\quad m_2=n_1-n_3,\quad m_3=\frac{1}{p}(n_1-n_2),\quad m_4=0.
\end{equation}
As $m_4=0$, we are left with a triplet of integers $\vec{m}$ corresponding to eigenvalues under the $T^3$-action on $Y^{p,q}$. Introducing the inward-pointing primitive normals of the toric diagram of $Y^{p,q}$
\begin{equation}
    \vec{v}_1=[1,0,0],\quad\vec{v}_2=[1,1,p],\quad\vec{v}_3=[1,2,p-q],\quad\vec{v}_4=[1,1,0],
\end{equation}
the triplet takes values in the dual moment map cone $\mathcal{C}_Y$
\begin{equation}\label{eq.CY}
    \mathcal{C}_Y=\{\vec{m}\in\mathbb{Z}^3\,|\, \vec{v}_1\cdot\vec{m}\geq 0,\; \vec{v}_2\cdot\vec{m}\geq 0,\; \vec{v}_3\cdot\vec{m}\geq 0,\; \vec{v}_4\cdot\vec{m}\geq 0\}.
\end{equation}
In \eqref{eq.conelattice} we also defined $\vec{\R}=A^{-1}\vec{\omega}$
\begin{equation}\label{eq.vecR}
    \R_1=\sum_{i=1}^4\omega_i,\quad\R_2=\omega_1+\omega_2-2\omega_4,\quad\R_3=p\omega_2+(p-q)\omega_4,\quad\R_4=\frac{1-ap}{p+1}\omega_2+\frac{2-a(p-q)}{p+q}\omega_4.
\end{equation}
We then arrive at the following rewriting\footnote{As the component $\R_4$ does not appear in the rewriting, with a slight abuse of notation we use $\vec{\R}$ to denote the vector with components $(\R_1,\R_2,\R_3)$.}
\begin{equation}\label{eq.1loop.red}
	Z^\text{vm}_{Y^{p,q}}|_{\mathfrak{m}_1=0}=\prod_{\alpha\in\Delta}\prod_{\vec{m}\in\mathcal{C}_Y}\left(\vec{m}\cdot\vec{\R}+\ii\alpha(a)\right)\prod_{\alpha\in\Delta}\prod_{\vec{m}\in\mathcal{C}_Y^\circ}\left(\vec{m}\cdot\vec{\R}-\ii\alpha(a)\right).
\end{equation}
A similar rewriting can be done for $L^{a,b,c}$ however, as we will not reduce these expressions to 4d, we will stick to present perturbative partition functions in terms of the lattice $\Lambda_L$.

Instanton contributions are obtained by gluing Nekrasov partition functions \cite{Nekrasov:2002qd,Nekrasov:2003rj} at each of the $m$ fixed fibers of $\widehat{M}_{\boldsymbol{p}}$. The changes to the Nekrasov partition function in presence of twist defects are presented in \cite{Nishioka:2016guu,Mauch:2025irx}. Their dependence on the flux sector will be discussed in \autoref{sec.5}.

\subsection{Perturbative Partition Function}\label{sec.3.2}
We now briefly recall the method employed in \cite{Qiu:2013pta} to compute the perturbative partition function \eqref{eq.1looproot} on $Y^{p,q}$ and $L^{a,b,c}$. This method, which relies on viewing these manifolds as quotients of $S^3\times S^3$, can be immediately generalized to compute the perturbative partition function also on the corresponding theories on $\widehat{Y}^{p,q}_{\boldsymbol{p}}$ and $\widehat{L}^{a,b,c}_{\boldsymbol{p}}$.

\paragraph{Dimensional Reduction from $S^3\times S^3$.}
Following standard arguments \cite{Pestun:2007rz}, the perturbative partition function of certain SQFTs can be equivalently computed in terms of the equivariant index of a cohomological complex obtained via localization. On $Y^{p,q}$ such complex can be uplifted to $S^3\times S^3$ as the extra $S^1_Y$ acts freely \cite{Qiu:2013pta}. Then, once the index in 6d is computed, the index on $Y^{p,q}$ is obtained keeping only modes with trivial representation under $S^1_Y$ \cite{Atiyah:1974}. The last step requires to translate the index into a one-loop determinant. We focus on the complex associated to a vector multiplet; the hypermultiplet case can be treated analogously. This procedure is equivalent to starting from an $\mathcal{N}=(1,0)$ theory on $S^3\times S^3$ and dimensionally reducing to 5d.

The complex on $S^3\times S^3$ splits into individual complexes of the two $S^3$ factors \cite{Qiu:2013pta}. The Cartan of the isometry group on $S^3\times S^3$ is $T^4=U(1)_A^2\times U(1)_B^2$ where the two factors act on each $S^3$ separately. We denote $s_1,s_2$ and $t_1,t_2$ the coordinates on the two copies of $U(1)^2$. The relevant index of each individual complex is given by
\begin{equation}\begin{split}\label{eq.index1}
    \iind_{U(1)_A^2}[\overline{\partial}]=&\sum_{n_1,n_2\geq 0}s_1^{-n_1}s_2^{-n_2}-\sum_{n_1,n_2< 0}s_1^{-n_1}s_2^{-n_2},\\
    \iind_{U(1)_B^2}[\overline{\partial}]=&\sum_{n_3,n_4\geq 0}t_1^{-n_3}t_2^{-n_4}-\sum_{n_3,n_4< 0}t_1^{-n_3}t_2^{-n_4}.
\end{split}\end{equation}
The index on $S^3\times S^3$ then follows
\begin{equation}\begin{split}\label{eq.index2}
    \iind_{T^4}[\overline{\partial}]=&\iind_{U(1)_A^2}[\overline{\partial}]\cdot\iind_{U(1)_B^2}[\overline{\partial}]\\
    =&\sum_{n_1,n_2\geq 0\atop n_3,n_4\geq 0}s_1^{-n_1}s_2^{-n_2}t_1^{-n_3}t_2^{-n_4}-\sum_{n_1,n_2\geq 0\atop n_3,n_4< 0}s_1^{-n_1}s_2^{-n_2}t_1^{-n_3}t_2^{-n_4}\\
    -&\sum_{n_1,n_2< 0\atop n_3,n_4\geq 0}s_1^{-n_1}s_2^{-n_2}t_1^{-n_3}t_2^{-n_4}+\sum_{n_1,n_2< 0\atop n_3,n_4< 0}s_1^{-n_1}s_2^{-n_2}t_1^{-n_3}t_2^{-n_4}.
\end{split}\end{equation}
The charge for rotation along $S^1_Y$ is \eqref{eq.chargesYL}
\begin{equation}\label{eq.chargesS1}
    u=(p+q)n_1+(p-q)n_2-pn_3-pn_4.
\end{equation}
Thus, keeping only the trivial representation and neglecting non-trivial KK modes, it is equivalent to impose
\begin{equation}\label{eq.u0}
    u=0\;\Rightarrow\;n_4=\frac{p+q}{p}n_1+\frac{p-q}{p}n_2-n_3.
\end{equation}
As $p>q>0$, the only terms in \eqref{eq.index2} which satisfy this condition are those with either all $n_i\geq 0$ or with all $n_i<0$.

Hence, the index on $Y^{p,q}$ is given by the quadruples $\vec{n}$ in \eqref{eq.index2} satisfying $u=0$. This defines a lattice
\begin{equation}\label{eq.LambdaY}
    \Lambda_{Y}=\{\vec{n}\in\mathbb{Z}^4_{\geq 0}\,|\,u=0\},
\end{equation}
and equivalently for negative $n_i$. Adding the character for the adjoint representation for the vector multiplet, and translating into a one-loop determinant, we find\footnote{In the second factor, we flipped the sign of the contributions in \eqref{eq.index2} with $n_i<0$ while also flipping the sign of $\ii\alpha(a)$.}
\begin{equation}\label{eq.1loop1}
    Z^\text{vm}_{Y^{p,q}}|_{\mathfrak{m}_1=0}=\prod_{\alpha\in\Delta}\prod_{\vec{n}\in\Lambda_{Y}}\left(\vec{n}\cdot\vec{\omega}+\ii\alpha(a)\right)\prod_{\alpha\in\Delta}\prod_{\vec{n}\in\Lambda^\circ_{Y}}\left(\vec{n}\cdot\vec{\omega}-\ii\alpha(a)\right).
\end{equation}
This expression reproduces the perturbative partition function in \eqref{eq.1looproot} with the lattice in \eqref{eq.LambdaM} given by the condition
\begin{equation}
    u=\vec{Q}^Y\cdot\vec{n}=0
\end{equation}
in \eqref{eq.LambdaY}. As explained above, we can use the matrix \eqref{eq.matrix} to rewrite the perturbative partition function as in \eqref{eq.1loop.red}. Along the same lines, one can also derive the perturbative partition function for the hypermultiplet \cite{Qiu:2013pta}. The main difference is an overall minus sign in \eqref{eq.index2} which leads to a term at the denominator in \eqref{eq.1looproot}.

As observed in \cite{Qiu:2013pta}, we can generalize to $L^{a,b,c}=(S^3\times S^3)/S^1_L$ this approach to compute the perturbative partition function. The corresponding charge for rotations along the free $S^1_L$ \eqref{eq.chargesYL} is given by
\begin{equation}
    \tilde{u}=an_1+bn_2-cn_3-(a+b-c)n_4,
\end{equation}
and the index of the relevant complex from localization is obtained from \eqref{eq.index2} imposing $\tilde{u}=0$. As $a,b,c>0$ and $a+b>c$, also in this case only the first and and last terms in \eqref{eq.index2} contribute. The one-loop determinant then is a product over a lattice
\begin{equation}
    \Lambda_{L^{a,b,c}}=\{\vec{n}\in\mathbb{Z}^4_{\geq 0}\,|\,\tilde{u}=0\}.
\end{equation}
Also in this case it possible to rewrite the perturbative partition function as a product over integers $\vec{m}$ in the dual moment map cone $\mathcal{C}_L$ of $L^{a,b,c}$.

\paragraph{Dimensional Reduction from Branched Cover of $S^3\times S^3$.}
If we now start instead from an $\mathcal{N}=(1,0)$ theory on $\widehat{S}^3_{\boldsymbol{p}_A}\times\widehat{S}^3_{\boldsymbol{p}_B}$, we can obtain the perturbative partition for an $\mathcal{N}=1$ theory on $\widehat{Y}^{p,q}_{\boldsymbol{p}}$ or $\widehat{L}^{a,b,c}_{\boldsymbol{p}}$. As in the smooth case, the perturbative partition function is given by the equivariant index which, as in \eqref{eq.index1}-\eqref{eq.index2}, can be computed taking the product of the individual indices on $\widehat{S}^3_{\boldsymbol{p}_A}$ and $\widehat{S}^3_{\boldsymbol{p}_B}$. On the two factors the index is given by that computed on a smooth $S^3$ with squashing parameters given by \cite{Nishioka:2013haa}
\begin{equation}
    \omega_A=(p^{-1}_1,p^{-1}_2),\qquad\omega_B=(p^{-1}_3,p^{-1}_4).
\end{equation}
Thus, we find
\begin{equation}\begin{split}\label{eq.index3}
    \iind_{U(1)_A^2}[\overline{\partial}]=&\sum_{n_1,n_2\geq 0}s_1^{-\frac{n_1}{p_1}}s_2^{-\frac{n_2}{p_2}}-\sum_{n_1,n_2< 0}s_1^{-\frac{n_1}{p_1}}s_2^{-\frac{n_2}{p_2}},\\
    \iind_{U(1)_B^2}[\overline{\partial}]=&\sum_{n_3,n_4\geq 0}t_1^{-\frac{n_3}{p_3}}t_2^{-\frac{n_4}{p_4}}-\sum_{n_3,n_4< 0}t_1^{-\frac{n_3}{p_3}}t_2^{-\frac{n_4}{p_4}},
\end{split}\end{equation}
and their product gives the index on $\widehat{S}^3_{\boldsymbol{p}_A}\times \widehat{S}^3_{\boldsymbol{p}_B}$
\begin{equation}\begin{split}\label{eq.index4}
    \iind_{T^4}[\overline{\partial}]=&\iind_{U(1)_A^2}[\overline{\partial}]\cdot\iind_{U(1)_B^2}[\overline{\partial}]\\
    =&\sum_{n_1,n_2\geq 0\atop n_3,n_4\geq 0}s_1^{-\frac{n_1}{p_1}}s_2^{-\frac{n_2}{p_2}}t_1^{-\frac{n_3}{p_3}}t_2^{-\frac{n_4}{p_4}}-\sum_{n_1,n_2\geq 0\atop n_3,n_4< 0}s_1^{-\frac{n_1}{p_1}}s_2^{-\frac{n_2}{p_2}}t_1^{-\frac{n_3}{p_3}}t_2^{-\frac{n_4}{p_4}}\\
    -&\sum_{n_1,n_2< 0\atop n_3,n_4\geq 0}s_1^{-\frac{n_1}{p_1}}s_2^{-\frac{n_2}{p_2}}t_1^{-\frac{n_3}{p_3}}t_2^{-\frac{n_4}{p_4}}+\sum_{n_1,n_2< 0\atop n_3,n_4< 0}s_1^{-\frac{n_1}{p_1}}s_2^{-\frac{n_2}{p_2}}t_1^{-\frac{n_3}{p_3}}t_2^{-\frac{n_4}{p_4}}.
\end{split}\end{equation}
The charges of the modes under $T^4$-rotations are multiples of $1/p_i$, consistently with a Fourier expansion on a $S^1$ whose periodicity is $2\pi p_i$. This behaviour is typical of branched covers, as for example on $S^5_{\boldsymbol{p}}$ \cite{Nishioka:2016guu}.

Focusing on $\widehat{Y}^{p,q}_{\boldsymbol{p}}$, we consider the locally free $S^1_{\widehat{Y}}$-action on $\widehat{S}^3_{\boldsymbol{p}_A}\times \widehat{S}^3_{\boldsymbol{p}_B}$ \eqref{eq.chargesYL.alpha} whose corresponding charges are
\begin{equation}\label{eq.ualpha}
    \frac{u_{\boldsymbol{p}}}{|\boldsymbol{p}|}=n_1\frac{p+q}{p_1}+n_2\frac{p-q}{p_2}-n_3\frac{p}{p_3}-n_4\frac{p}{p_4}.
\end{equation}
The index on $\widehat{Y}^{p,q}_{\boldsymbol{p}}=\widehat{S}^3_{\boldsymbol{p}_A}\times \widehat{S}^3_{\boldsymbol{p}_B}/S^1_{\widehat{Y}}$ is obtained from \eqref{eq.index4} restricting to contributions with $u_{\boldsymbol{p}}=0$. Translating into a one-loop determinant, we find that the perturbative partition function is given by
\begin{equation}\label{eq.1loop2}
    Z^\text{vm}_{\widehat{Y}^{p,q}_{\boldsymbol{p}}}|_{\mathfrak{m}_1=0}=\prod_{\alpha\in\Delta}\prod_{\vec{n}\in\Lambda_{\widehat{Y}}}\left(\vec{n}_{\boldsymbol{p}}\cdot\vec{\omega}+\ii\alpha(a)\right)\prod_{\vec{n}\in\Lambda^\circ_{\widehat{Y}}}
    \left(\vec{n}_{\boldsymbol{p}}\cdot\vec{\omega}-\ii\alpha(a)\right),
\end{equation}
where
\begin{equation}
    \vec{n}_{\boldsymbol{p}}=\left(\frac{n_1}{p_1},\dots,\frac{n_4}{p_4}\right),
\end{equation}
and 
\begin{equation}
    \Lambda_{\widehat{Y}}=\{\vec{n}\in\mathbb{Z}^4_{\geq 0}\,|\,u_{\boldsymbol{p}}=0\}.
\end{equation}

The perturbative partition function for the hypermultiplet is obtained keeping track of the overall minus sign in \eqref{eq.index4}, the dependence on the mass $\hat{m}$ in \eqref{eq.1looproot} and on the constant shift by $\sum_{i=1}^4\frac{\omega_i}{p_i}\equiv\bar{\omega}_{\boldsymbol{p}}$. We find
\begin{equation}\begin{split}
    Z^\text{hm}_{\widehat{Y}^{p,q}_{\boldsymbol{p}}}|_{\mathfrak{m}=0}=\prod_{\rho\in R}&\prod_{\vec{n}\in\Lambda_{\widehat{Y}}}\left(\vec{n}_{\boldsymbol{p}}\cdot\vec{\omega}+\ii\hat{m}+\frac{1}{2}\bar{\omega}_{\boldsymbol{p}}+\ii\rho(a)\right)^{-1}\\
    &\prod_{\vec{n}\in\Lambda^\circ_{\widehat{Y}}}
    \left(\vec{n}_{\boldsymbol{p}}\cdot\vec{\omega}+\ii\hat{m}+\frac{1}{2}\bar{\omega}_{\boldsymbol{p}}-\ii\rho(a)\right)^{-1},
\end{split}\end{equation}
where the infinite products, as expected, appear at the denominator.

As in \eqref{eq.1loop.red}, we can use the matrix $A$ \eqref{eq.matrix} to write these expression in terms of charges under the $T^3$-rotation of $\widehat{Y}^{p,q}_{\boldsymbol{p}}$. We find that the relation between $\vec{n}$ and $\vec{m}$ is
\begin{equation}
    \frac{m_1}{p_3}=\frac{n_3}{p_3},\quad \frac{m_2}{p_1p_3}=\frac{n_3p_1-n_1p_3}{p_1p_3},\quad m_3=\frac{1}{p}\left(\frac{n_1p_2-n_2p_1}{p_1p_2}\right),\quad m_4=0
\end{equation}
This allows us to write
\begin{equation}\label{eq.1loop.red.alpha}
	Z^\text{vm}_{\widehat{Y}^{p,q}_{\boldsymbol{p}}}|_{\mathfrak{m}_1=0}=\prod_{\alpha\in\Delta}\prod_{\vec{m}\in\mathcal{C}_{\widehat{Y}}}\left(\vec{m}_{\boldsymbol{p}}\cdot\vec{\R}+\ii\alpha(a)\right)\prod_{\alpha\in\Delta}\prod_{\vec{m}\in\mathcal{C}^\circ_{\widehat{Y}}}\left(\vec{m}_{\boldsymbol{p}}\cdot\vec{\R}-\ii\alpha(a)\right),
\end{equation}
where  
\begin{equation}
    \vec{m}_{\boldsymbol{p}}=\left(\frac{m_1}{p_3},\frac{m_2}{p_1p_3},\frac{m_3}{p_1p_2}\right),
\end{equation}
$\mathcal{C}_{\widehat{Y}}$ is found to be equivalent to \eqref{eq.CY}, and $\vec{\R}$ is as in \eqref{eq.vecR}.

The same procedure applies to $\widehat{L}^{a,b,c}_{\boldsymbol{p}}$, simply quotienting $S^3_{\boldsymbol{p}_A}\times S^3_{\boldsymbol{p}_B}$ by the $S^1_{\widehat{L}}$-action in \eqref{eq.chargesYL.alpha} with charges
\begin{equation}
    \frac{\tilde{u}_{\boldsymbol{p}}}{|\boldsymbol{p}|}=n_1\frac{a}{p_1}+n_2\frac{b}{p_2}-n_3\frac{c}{p_3}-n_4\frac{a+b-c}{p_4}.
\end{equation}
The perturbative partition function is a product over the following lattice
\begin{equation}
    \Lambda_{\widehat{L}}=\{\vec{n}\in\mathbb{Z}^4_{\geq 0}\,|\,\tilde{u}_{\boldsymbol{p}}=0\}.
\end{equation}
As for the smooth case, it is not needed to express these expressions in terms of the dual moment map cone $\mathcal{C}_{\widehat{L}}$ as they cannot be reduced to 4d.

\section{Fluxes Via Dimensional Reduction}\label{sec.4}
In this section we present the main results of our work. First, focusing on the trivial instanton sector in  \eqref{eq.Zfull.M}, we compute the one-loop determinant at all flux sectors on $\widehat{Y}^{p,q}_{\boldsymbol{p}}$ and $\widehat{L}^{a,b,c}_{\boldsymbol{p}}$. Along the lines of \cite{Lundin:2021zeb,Lundin:2023tzw,Mauch:2024uyt}, we obtain these contributions keeping non-trivial representations under the free $S^1_{\widehat{Y}}$- and $S^1_{\widehat{L}}$-action when dimensionally reducing from $\widehat{S}^3_{\boldsymbol{p}_A}\times \widehat{S}^3_{\boldsymbol{p}_B}$. Without twist defects, that is for $\boldsymbol{p}=(1,1,1,1)$, our results improve those for the trivial flux sector derived in \cite{Qiu:2013pta} and presented in the previous section.

In the second part, we further dimensionally reduce the $\mathcal{N}=1$ theory to an $\mathcal{N}=2$ theory on $\widehat{B}_{\boldsymbol{p}}=\widehat{Y}^{p,q}_{\boldsymbol{p}}/S^1$. This will enable us to compute the dependence, of the one-loop determinant, on fluxes through both independent two-cycles of $\widehat{B}_{\boldsymbol{p}}$. For theories without twist defects, this improves the result for the one-loop determinant computed in \cite{Lundin:2023tzw} for fluxes through only one two-cycle.

According to the proposal in \cite{Mauch:2025irx}, the partition functions we compute in this section for theories with twist defects, for vector multiplets and hypermultiplets in the adjoint representation, reproduce the partition functions on spaces with orbifold singularities discussed in \autoref{sec.2}.

\subsection{Dimensional Reduction}
The method we use to compute one-loop determinants at each flux sector relies on a dimensional reduction along an at least locally free non-trivial $S^1$-fiber
\begin{equation}
    S^1\hookrightarrow \widehat{X}^{d+1}_{\boldsymbol{p}}\rightarrow \widehat{X}^d_{\boldsymbol{p}},
\end{equation}
The first example of such dimensional reduction appeared in \cite{Benini:2012ui}, where the one-loop determinant at all flux sectors of an $\mathcal{N}=(2,2)$ theory on $\mathbb{CP}^1$ has been reproduced from the partition function of an $\mathcal{N}=2$ theory on lens spaces $L(h,-1)$ \cite{Alday:2012au}, taking the large $h$ limit. More recently, this method has been thoroughly analysed in \cite{Lundin:2021zeb,Lundin:2023tzw}, where it has been applied to the dimensional reduction of gauge theories on toric Sasakian manifolds down to quasi-toric four-manifolds. Moreover, it has been applied to compute flux contributions on weighted projective spaces $\mathbb{CP}^r_{\boldsymbol{p}}$ dimensionally reducing from branched covers $\widehat{S}^{2r-1}_{\boldsymbol{p}}$ \cite{Mauch:2024uyt}.

Let us briefly summarize the mechanism through which fluxes arise via dimensional reduction; a detailed exposition can be found in \cite{Lundin:2023tzw}. The starting point is the perturbative partition function of an SQFT on $\widehat{X}^{d+1}_{\boldsymbol{p}}$, which depends on the charges $\vec{n}$ of the modes under the torus action on $\widehat{X}^{d+1}_{\boldsymbol{p}}$ (see e.g. \eqref{eq.1loop.red}). We can rewrite such object in terms of the charge for rotations along the free $S^1$-fiber, $t=f(\vec{n})$, where $f(\cdot)$ is a linear function with integer coefficients. 

Rather than shrinking the radius of the $S^1$-fiber, we introduce a quotient space\footnote{Note that for locally free $S^1$-actions, as \eqref{eq.chargesYL.alpha} on the branched cover of $S^3\times S^3$, we take $\gcd(h,p_i)=1,\,\forall i$. In this way, the $\mathbb{Z}_h$-action is actually globally free. The final result however is not affected by this choice. Indeed, the effect of quotenting by $\mathbb{Z}_{p_i}$ is that of identifying $p_i$ sheets \cite{Albach_2021} without affecting the first fundamental group. After a quotient by $\mathbb{Z}_{|\boldsymbol{p}|}$ one finds the smooth manifold $S^3\times S^3$. However, the theory on $S^3\times S^3$ includes twist defects \cite{Calabrese:2009qy}, and thus it is not the standard theory on $S^3\times S^3$. This theory is equivalent to the original theory on the branched cover of $S^3\times S^3$, as shown for spheres in \cite{Nishioka:2016guu}. On $S^3\times S^3$ the quotient by $\mathbb{Z}_h$ is globally free $\forall\,h$, and taking the large $h$ limit is equivalent to take the large $h$ limit of $(\widehat{S}^3_{\boldsymbol{p}_A}\times\widehat{S}^3_{\boldsymbol{p}_B})/\mathbb{Z}_h$ with $\gcd(h,p_i)=1$.} $\widehat{X}^{d+1}_{\boldsymbol{p}}/\mathbb{Z}_h$ with $\pi_1(\widehat{X}^{d+1}_{\boldsymbol{p}}/\mathbb{Z}_h)=\mathbb{Z}_h$. To each element in $\pi_1$ there exists a non-trivial line bundle and an associated flat gauge connection over which one needs to sum in the partition function. At each topological sector, the SQFT contains only modes satisfying the projection condition
\begin{equation}\label{eq.projection.quotient}
    t=\alpha(\mathfrak{m})\mmod h,\qquad\mathfrak{m}=\text{diag}(n_1,\dots,m_{\text{rk }G}),\quad 0\leq n_i<h-1,
\end{equation}
where $\mathfrak{m}$ are the winding numbers of the flat gauge connection. Therefore, the one-loop determinant at each topological sector is found by imposing \eqref{eq.projection.quotient} on the perturbative partition function on $\widehat{X}^{d+1}_{\boldsymbol{p}}$. 

At large $h$, we obtain the space $\widehat{X}^d_{\boldsymbol{p}}=\widehat{X}^{d+1}_{\boldsymbol{p}}/S^1$, where $H_2(\widehat{X}^d_{\boldsymbol{p}},\mathbb{Z})\simeq H_2(\widehat{X}^{d+1}_{\boldsymbol{p}},\mathbb{Z})\times \mathbb{Z}$. The sum over flat connections in the partition function gives rise to a sum over fluxes $\mathfrak{m}$ labelled by the first Chern class of the line bundle. The one-loop determinant in each flux sector is obtained restricting to 
\begin{equation}\label{eq.projection.base}
    t=\alpha(\mathfrak{m})
\end{equation}
in the perturbative partition function on $\widehat{X}^{d+1}_{\boldsymbol{p}}$. Thus, unlike standard Kaluza-Klein dimensional reduction, when reducing along a non-trivial $S^1$-fiber all KK modes are present in the $d$-dimensional theory, and distribute into the new topological sectors on the base space. 

The dimensional reduction in \autoref{sec.3}, which we employed to compute the perturbative partition function on $\widehat{Y}^{p,q}_{\boldsymbol{p}}$, is along a non-trivial $S^1$-fiber. There, we restricted the equivariant index on $\widehat{S}^3_{\boldsymbol{p}_A}\times \widehat{S}^3_{\boldsymbol{p}_B}$ to the trivial charge under the $S^1_{\widehat{Y}}$-action \eqref{eq.ualpha}. Therefore, keeping also non-trivial charges, we will find the one-loop determinant at all flux sectors on the base space. We simply have to restrict the equivariant index in \eqref{eq.index4} to
\begin{equation}\begin{split}
    \widehat{Y}^{p,q}_{\boldsymbol{p}}:\quad u_{\boldsymbol{p}}=&\alpha(\mathfrak{m}_1),\\
    \widehat{L}^{a,b,c}_{\boldsymbol{p}}:\quad \tilde{u}_{\boldsymbol{p}}=&\alpha(\mathfrak{m}_1).
\end{split}\end{equation}
For what concerns the classical contribution at a given flux sector, we will assume that the expression at the trivial flux sector \eqref{eq.classical.M} continues to hold. The same is shown to be true dimensionally reducing from 5d to 4d \cite{Lundin:2021zeb,Lundin:2023tzw}. In those cases, a flux dependence of the classical contributions can be achieved via a shift of the Coulomb branch parameter.

Furthermore, starting this time from the one-loop determinant at all flux sectors on $\widehat{Y}^{p,q}_{\boldsymbol{p}}$, we can compute the one-loop determinant of an $\mathcal{N}=2$ theory on $\widehat{B}_{\boldsymbol{p}}$. This result will contain contributions from two fluxes as $H_2(\widehat{B}_{\boldsymbol{p}},\mathbb{Z})\simeq\mathbb{Z}^2$, and it is obtained simply by restricting the equivariant index \eqref{eq.index4} to contributions with
\begin{equation}
    u_{\boldsymbol{p}}=\alpha(\mathfrak{m}_1),\qquad t_{\boldsymbol{p}}=\alpha(\mathfrak{m}_2).
\end{equation}
On these 4d spaces, the partition function is given by
\begin{equation}\label{eq.Zfull.B}
    \mathcal{Z}_{\widehat{B}_{\boldsymbol{p}}}=\sum_{\mathfrak{m}_1,\mathfrak{m}_2}\int_{\mathfrak{h}}\dd a\, e^{-S_\text{cl}}\cdot Z_{\widehat{B}_{\boldsymbol{p}}}\cdot Z_{\widehat{B}_{\boldsymbol{p}}}^\text{inst}.
\end{equation}
As mentioned above, the classical action is not affect by the flux sector. Thus, we find\footnote{See \cite{Lundin:2023tzw} for how to implement the limit of large $h$ at the level of the classical contribution.} 
\begin{equation}
    S_\text{cl}=-\frac{4\pi^2}{g^2_\text{YM,4d}(x)}\varrho_\text{4d}\,\text{tr}(a^2),
\end{equation}
where $\varrho|_\text{4d}$ is a function of $\vec{\omega}$ and the coupling constant depends on the position\footnote{This is due to the fiber employed in dimensional reduction not being the Reeb vector whenever the toric Sasakian space is not regular.} $x\in \widehat{B}_{\boldsymbol{p}}$ \cite{Festuccia:2016gul}. Let us recall that the theory we find in 4d is an exotic theory \cite{Festuccia:2018rew}, a generalization of Pestun's theory to generic four manifolds, and with field strength localizing to instantons and anti-instantons at different fixed points\footnote{In \cite{Festuccia:2018rew} it is shown how exotic, or Pestun-like, theories and topologically twisted theories on four dimensional compact manifolds arise as two instances of a unique framework, named \emph{Pestunization}. The idea is that given a 4d $\mathcal{N}=2$ SYM with a $T^2$-isometry and a choice of assigning ``$+$'' or ``$-$'' to each chart of spacetime containing a fixed point, the output of Pestunization is an equivariant cohomological field theory, whose BPS field strength is self-dual on the ``$+$'' charts and anti-delf-dual on the ``$-$'' charts. Topologically twisted theory is the special case where one assigns ``$+$'' to all charts while in exotic theories the field strength flips from ``$+$'' to ``$-$'' at different fixed points.}.

\subsection{6d to 5d}
We start our journey through dimensional reductions in 6d. Here, we will show how to use the equivariant index on $\widehat{S}^3_{\boldsymbol{p}_A}\times \widehat{S}^3_{\boldsymbol{p}_B}$ to compute the one-loop determinant, at all flux sectors, on $\widehat{Y}^{p,q}_{\boldsymbol{p}}$ and $\widehat{L}^{a,b,c}_{\boldsymbol{p}}$. We will first deal with the theories without twist defects and later consider theories where they are present.

\paragraph{Dimensional Reduction from $S^3\times S^3$.}
The quotients of $S^3\times S^3$ by the free $S^1$-actions \eqref{eq.chargesYL}, lead to $Y^{p,q}$ and $L^{a,b,c}$. Focusing, first, on $Y^{p,q}$, we need to restrict the equivariant index \eqref{eq.index2} to charges $\vec{n}$ satisfying
\begin{equation}
    u=(p+q)n_1+(p-q)n_2-pn_3-pn_4=\alpha(\mathfrak{m}_1).
\end{equation}
For $\alpha(\mathfrak{m}_1)=0$ we argued, in the previous section, that only contributions with all $n_i\geq 0$ or all $n_i<0$ contribute. For generic $\alpha(\mathfrak{m})\neq 0$, two new phenomena occur.
\begin{itemize}
    \item Terms in \eqref{eq.index2} with $n_1,n_2\geq 0$ and $n_3,n_4<0$ (or $n_1,n_2< 0$ and $n_3,n_4\geq 0$) contribute. These come with a minus sign in the index and will appear at the denominator for a vector multiplet. Instead, they will be at the numerator for a hypermultiplet.
    \item There is no symmetry $\vec{n}\rightarrow -\vec{n}$ among the terms in \eqref{eq.index2} contributing. To reinstate such symmetry, we would have to restrict the product over positive roots.
\end{itemize}
Following this discussion, we define the following lattices
\begin{equation}\begin{split}\label{eq.Lambda.n}
    \Lambda^{++}_{Y}(\mathfrak{m}_1)=&\{\vec{n}\in\mathbb{Z}^4_{\geq 0}\,|\,u=\alpha(\mathfrak{m}_1)\},\\
    \Lambda^{+-}_{Y}(\mathfrak{m}_1)=&\{(n_1,n_2)\in\mathbb{Z}^2_{\geq 0},\,(n_3,n_4)\in\mathbb{Z}^2_{< 0}\,|\,u=\alpha(\mathfrak{m}_1)\},\\
    \Lambda^{-+}_{Y}(\mathfrak{m}_1)=&\{(n_1,n_2)\in\mathbb{Z}^2_{< 0},\,(n_3,n_4)\in\mathbb{Z}^2_{\geq 0}\,|\,u=\alpha(\mathfrak{m}_1)\},\\
    \Lambda^{--}_{Y}(\mathfrak{m}_1)=&\{\vec{n}\in\mathbb{Z}^4_{< 0}\,|\,u=\alpha(\mathfrak{m}_1)\}.
\end{split}\end{equation}
Correspondingly, for a vector multiplet the one-loop determinant at a flux sector is
\begin{equation}\label{eq.1loop.n}
     Z^\text{vm}_{Y^{p,q}}=\prod_{\alpha\in\Delta}\frac{\prod\limits_{\vec{n}\in\Lambda^{++}_{Y}(\mathfrak{m}_1)}\left(\vec{n}\cdot\vec{\omega}+\ii\alpha(a)\right)\prod\limits_{\vec{n}\in\Lambda^{--}_{Y}(\mathfrak{m}_1)}\left(\vec{n}\cdot\vec{\omega}+\ii\alpha(a)\right)}{\prod\limits_{\vec{n}\in\Lambda^{+-}_{Y}(\mathfrak{m}_1)}\left(\vec{n}\cdot\vec{\omega}+\ii\alpha(a)\right)\prod\limits_{\vec{n}\in\Lambda^{-+}_{Y}(\mathfrak{m}_1)}\left(\vec{n}\cdot\vec{\omega}+\ii\alpha(a)\right)}.
\end{equation}
For a hypermultiplet we find
\begin{equation}
    Z^\text{hm}_{Y^{p,q}}=\prod_{\rho\in R}\frac{\prod\limits_{\vec{n}\in\Lambda^{+-}_{Y}(\mathfrak{m}_1)}\left(\vec{n}\cdot\vec{\omega}+\ii\hat{m}+\frac{1}{2}\bar{\omega}+\ii\rho(a)\right)\prod\limits_{\vec{n}\in\Lambda^{-+}_{Y}(\mathfrak{m}_1)}\left(\vec{n}\cdot\vec{\omega}+\ii\hat{m}+\frac{1}{2}\bar{\omega}+\ii\rho(a)\right)}{\prod\limits_{\vec{n}\in\Lambda^{++}_{Y}(\mathfrak{m}_1)}\left(\vec{n}\cdot\vec{\omega}+\ii\hat{m}+\frac{1}{2}\bar{\omega}+\ii\rho(a)\right)\prod\limits_{\vec{n}\in\Lambda^{--}_{Y}(\mathfrak{m}_1)}\left(\vec{n}\cdot\vec{\omega}+\ii\hat{m}+\frac{1}{2}\bar{\omega}+\ii\rho(a)\right)}.
\end{equation}
These expressions generalize the results of \cite{Qiu:2013aga,Qiu:2013pta} computed at trivial flux sector.

As for the trivial flux \eqref{eq.1loop.red}, we can use the matrix $A$ \eqref{eq.matrix} to rewrite these one-loop determinants in terms of the triplet of integers $\vec{m}$ for $T^3$-rotations in $Y^{p,q}$. As we are considering $u\neq 0$, unlike in \eqref{eq.conelattice}, we find
\begin{equation}\begin{split}
    \prod_{\vec{n}\in\Lambda_Y}(\vec{n}\cdot\vec{\omega}+x)=&\prod_{\vec{m}\in\mathcal{C}_Y}\left(\vec{m}\cdot\vec{\R}+x+\left(\R_3\frac{1-ap}{p(p+q)}-\R_4\right)\alpha(\mathfrak{m}_1)\right)\\
    =&\prod_{\vec{m}\in\mathcal{C}_Y}\left(\vec{m}\cdot\vec{\R}+x-\frac{\omega_4}{p}\alpha(\mathfrak{m}_1)\right)
\end{split}\end{equation}
where $\vec{m}$ is as in \eqref{eq.m_i}, and in the second line we employed the definitions of $\R_3,\R_4$ in terms of $\omega_i$ to simplify the expression. Taking into account these modifications, we find
\begin{equation}\begin{split}\label{eq.1loop.red.n}
	Z^\text{vm}_{Y^{p,q}}=\prod_{\alpha\in\Delta}&\frac{\prod\limits_{\vec{m}\in\mathcal{C}^{++}_{Y}(\mathfrak{m}_1)}\left(\vec{m}\cdot\vec{\R}+\ii\alpha(a)-\frac{\omega_4}{p}\alpha(\mathfrak{m}_1)\right)\prod\limits_{\vec{m}\in\mathcal{C}^{--}_{Y}(\mathfrak{m}_1)}\left(\vec{m}\cdot\vec{\R}+\ii\alpha(a)-\frac{\omega_4}{p}\alpha(\mathfrak{m}_1)\right)}{\prod\limits_{\vec{m}\in\mathcal{C}^{+-}_{Y}(\mathfrak{m}_1)}\left(\vec{m}\cdot\vec{\R}+\ii\alpha(a)-\frac{\omega_4}{p}\alpha(\mathfrak{m}_1)\right)\prod\limits_{\vec{m}\in\mathcal{C}^{-+}_{Y}(\mathfrak{m}_1)}\left(\vec{m}\cdot\vec{\R}+\ii\alpha(a)-\frac{\omega_4}{p}\alpha(\mathfrak{m}_1)\right)}.
\end{split}\end{equation}
The product is over the lattices
\begin{equation}\begin{split}\label{eq.C.n}
	\mathcal{C}^{++}_{Y}(\mathfrak{m}_1)=&\left\{\vec{m}\in\mathbb{Z}^3\,|\, \vec{v}_1\cdot\vec{m}\geq 0,\; \vec{v}_2\cdot\vec{m}\geq 0,\; \vec{v}_3\cdot\vec{m}\geq \frac{\alpha(\mathfrak{m}_1)}{p},\; \vec{v}_4\cdot\vec{m}\geq 0\right\},\\
	\mathcal{C}^{+-}_{Y}(\mathfrak{m}_1)=&\left\{\vec{m}\in\mathbb{Z}^3\,|\, \vec{v}_1\cdot\vec{m}\leq 0,\; \vec{v}_2\cdot\vec{m}\geq 0,\; \vec{v}_3\cdot\vec{m}\leq \frac{\alpha(\mathfrak{m}_1)}{p},\; \vec{v}_4\cdot\vec{m}\geq 0\right\},\\
	\mathcal{C}^{-+}_{Y}(\mathfrak{m}_1)=&\left\{\vec{m}\in\mathbb{Z}^3\,|\, \vec{v}_1\cdot\vec{m}\geq 0,\; \vec{v}_2\cdot\vec{m}\leq 0,\; \vec{v}_3\cdot\vec{m}\geq \frac{\alpha(\mathfrak{m}_1)}{p},\; \vec{v}_4\cdot\vec{m}\leq0\right\},\\
	\mathcal{C}^{--}_{Y}(\mathfrak{m}_1)=&\left\{\vec{m}\in\mathbb{Z}^3\,|\, \vec{v}_1\cdot\vec{m}\leq 0,\; \vec{v}_2\cdot\vec{m}\leq 0,\; \vec{v}_3\cdot\vec{m}\leq \frac{\alpha(\mathfrak{m}_1)}{p},\; \vec{v}_4\cdot\vec{m}\leq0\right\}.\\
\end{split}\end{equation}
As usual, one can derive the same expression also for hypermultiplets. 

Similar results on $L^{a,b,c}$ are obtained straightforwardly simply employing the corresponding $S^1_L$-action in \eqref{eq.chargesYL}.

\paragraph{Dimensional Reduction from Branched Cover of $S^3\times S^3$.}
To compute the one-loop determinant at all flux sectors on $\widehat{Y}^{p,q}_{\boldsymbol{p}}$ we simply have to employ the charges for rotations along the $S^1_{\widehat{Y}}$-action in \eqref{eq.chargesYL.alpha}
\begin{equation}
    \frac{u_{\boldsymbol{p}}}{|\boldsymbol{p}|}=n_1\frac{p+q}{p_1}+n_2\frac{p-q}{p_2}-n_3\frac{p}{p_3}-n_4\frac{p}{p_4}.
\end{equation}
Then, in analogy with \eqref{eq.Lambda.n}, we define the following lattices
\begin{equation}\begin{split}\label{eq.Lambda.alpha.n}
    \Lambda^{++}_{\widehat{Y}}(\mathfrak{m}_1)=&\{\vec{n}\in\mathbb{Z}^4_{\geq 0}\,|\,\frac{u_{\boldsymbol{p}}}{|\boldsymbol{p}|}=\alpha(\mathfrak{m}_1)\},\\
    \Lambda^{+-}_{\widehat{Y}}(\mathfrak{m}_1)=&\{(n_1,n_2)\in\mathbb{Z}^2_{\geq 0},\,(n_3,n_4)\in\mathbb{Z}^2_{< 0}\,|\,\frac{u_{\boldsymbol{p}}}{|\boldsymbol{p}|}=\alpha(\mathfrak{m}_1)\},\\
    \Lambda^{-+}_{\widehat{Y}}(\mathfrak{m}_1)=&\{(n_1,n_2)\in\mathbb{Z}^2_{< 0},\,(n_3,n_4)\in\mathbb{Z}^2_{\geq 0}\,|\,\frac{u_{\boldsymbol{p}}}{|\boldsymbol{p}|}=\alpha(\mathfrak{m}_1)\},\\
    \Lambda^{--}_{\widehat{Y}}(\mathfrak{m}_1)=&\{\vec{n}\in\mathbb{Z}^4_{< 0}\,|\,\frac{u_{\boldsymbol{p}}}{|\boldsymbol{p}|}=\alpha(\mathfrak{m}_1)\},
\end{split}\end{equation}
which contribute to the one-loop determinant
\begin{equation}\label{eq.1loop.alpha.n}
     Z^\text{vm}_{\widehat{Y}^{p,q}_{\boldsymbol{p}}}=\prod_{\alpha\in\Delta}\frac{\prod\limits_{\vec{n}\in\Lambda^{++}_{\widehat{Y}}(\mathfrak{m}_1)}\left(\vec{n}\cdot\vec{\omega}+\ii\alpha(a)\right)\prod\limits_{\vec{n}\in\Lambda^{--}_{\widehat{Y}}(\mathfrak{m}_1)}\left(\vec{n}\cdot\vec{\omega}+\ii\alpha(a)\right)}{\prod\limits_{\vec{n}\in\Lambda^{+-}_{\widehat{Y}}(\mathfrak{m}_1)}\left(\vec{n}\cdot\vec{\omega}+\ii\alpha(a)\right)\prod\limits_{\vec{n}\in\Lambda^{-+}_{\widehat{Y}}(\mathfrak{m}_1)}\left(\vec{n}\cdot\vec{\omega}+\ii\alpha(a)\right)}.
\end{equation}
As in \eqref{eq.1loop.red.n} we can use the matrix $A$ to find
\begin{equation}\label{eq.1loop.red.n.alpha}
	Z^\text{vm}_{Y^{p,q}}=\prod_{\alpha\in\Delta}\frac{\prod\limits_{\vec{m}\in\mathcal{C}^{++}_{\widehat{Y}}(\mathfrak{m}_1)}\left(\vec{m}_{\boldsymbol{p}}\cdot\vec{\R}+\ii\alpha(a)-\frac{\omega_4}{p}\frac{\alpha(\mathfrak{m}_1)}{|\boldsymbol{p}|}\right)\prod\limits_{\vec{m}\in\mathcal{C}^{--}_{\widehat{Y}}(\mathfrak{m}_1)}\left(\vec{m}_{\boldsymbol{p}}\cdot\vec{\R}+\ii\alpha(a)-\frac{\omega_4}{p}\frac{\alpha(\mathfrak{m}_1)}{|\boldsymbol{p}|}\right)}{\prod\limits_{\vec{m}\in\mathcal{C}^{+-}_{\widehat{Y}}(\mathfrak{m}_1)}\left(\vec{m}_{\boldsymbol{p}}\cdot\vec{\R}+\ii\alpha(a)-\frac{\omega_4}{p}\frac{\alpha(\mathfrak{m}_1)}{|\boldsymbol{p}|}\right)\prod\limits_{\vec{m}\in\mathcal{C}^{-+}_{\widehat{Y}}(\mathfrak{m}_1)}\left(\vec{m}_{\boldsymbol{p}}\cdot\vec{\R}+\ii\alpha(a)-\frac{\omega_4}{p}\frac{\alpha(\mathfrak{m}_1)}{|\boldsymbol{p}|})\right)}
\end{equation}
where the lattices $\mathcal{C}^{\pm\pm}_{\widehat{Y}}$ are given by \eqref{eq.C.n} replacing $\alpha(\mathfrak{m}_1)$ with $\alpha(\mathfrak{m}_1)/|\boldsymbol{p}|$. Finally, expressions on $\widehat{L}^{a,b,c}_{\boldsymbol{p}}$ and for hypermultiplets follow straightforwardly.

\subsection{5d to 4d}
As discussed in \autoref{sec.2}, $L^{a,b,c}$ does not admit a free $S^1$-action. Hence we focus on $Y^{p,q}$ and the $S^1$-action in \eqref{eq.freeS1Y}, whose corresponding charge is
\begin{equation}
    t^\text{ex}=m_3.
\end{equation}
Similarly, for $\widehat{Y}^{p,q}_{\boldsymbol{p}}$, the $S^1$-action is that in \eqref{eq.freeS1Y}. However in this case, due to the possibility of fractional $m_{p,3}$, we need to modify the expression above as follows
\begin{equation}
    \frac{t^\text{ex}_{\boldsymbol{p}}}{p_1p_2}=\frac{m_3}{p_1p_2}=m_{\boldsymbol{p},3}.
\end{equation}
Recall that this choice of $t$ leads to an exotic (or Pestun-like) theory \cite{Festuccia:2018rew}.

We will only discuss the case with twist defects. To recover the setup without their insertion it is enough to set all $p_i$ to one. We take, as starting point, the expressions written explicitly in terms for the $T^3$-action on $\widehat{Y}^{p,q}_{\boldsymbol{p}}$: the perturbative partition function \eqref{eq.1loop.red.alpha} and the one-loop determinant at a flux sector $\mathfrak{m}_1$ \eqref{eq.1loop.red.n.alpha}. Contributions at a given flux sector $\mathfrak{m}_2$ are obtained restricting to charges satisfying
\begin{equation}
    t_{\boldsymbol{p}}=\alpha(\mathfrak{m}_2).
\end{equation}
Let us start analysing the perturbative partition function in 5d \eqref{eq.1loop.red.alpha}. Substituting for $m_3$ we find
\begin{equation}\begin{split}\label{eq.1loop4.ex}
    Z^\text{vm,ex}_{\widehat{B}_{\boldsymbol{p}}}|_{\mathfrak{m}_1=0}=\prod_{\alpha\in\Delta}&\prod_{(m_1,m_2)\in\mathcal{B}_{\widehat{B}}(\mathfrak{m}_2)}\left(m_1\frac{\epsilon_1}{p_3}+m_2\frac{\epsilon_2}{p_1p_3}+\ii\alpha(a)+\frac{\R_3}{p_1p_2}\alpha(\mathfrak{m}_2)\right)\\
    &\prod_{(m_1,m_2)\in\mathcal{B}^{\circ}_{\widehat{B}}(\mathfrak{m}_2)}\left(m_1\frac{\epsilon_1}{p_3}+m_2\frac{\epsilon_2}{p_1p_3}-\ii\alpha(a)+\frac{\R_3}{p_1p_2}\alpha(\mathfrak{m}_2)\right),
\end{split}\end{equation}
where, we have redefined $\epsilon_1=\R_1$ and $\epsilon_2=\R_2$ to conform with notation for equivariant parameters in 4d. The lattices are given by
\begin{equation}\label{eq.Btopex}
    \mathcal{B}_{\widehat{B}}(\mathfrak{m}_2)=\text{proj}_{12}\left\{\vec{m}\in\mathcal{C}_{\widehat{Y}}\,|\,m_3=t_{\boldsymbol{p}}\right\}.
\end{equation}
where $\text{proj}_{12}[m_1,m_2,m_3]=[m_1,m_2]$. It is possible to write \eqref{eq.1loop4.ex} in terms of generalized $\Upsilon$-functions \cite{Hama:2012bg,Festuccia:2018rew,Lundin:2023tzw,Mauch:2024uyt}. For vanishing branching indices $p_i=1$, these expressions reproduce those found in \cite{Lundin:2023tzw}.

Moving to the one-loop around fluxes on $\widehat{Y}^{p,q}_{\boldsymbol{p}}$ \eqref{eq.1loop.red.n.alpha}, upon dimensional reduction we find
\begin{equation}\begin{split}\label{eq.1loop4.ex.fin}
    Z^\text{vm,ex}_{\widehat{B}_{\boldsymbol{p}}}=\prod_{\alpha\in\Delta}&\frac{\prod\limits_{(m_1,m_2)\in\mathcal{B}^{++}_{\widehat{B}}(\mathfrak{m}_1,\mathfrak{m}_2)}\left(m_1\frac{\epsilon_1}{p_3}+m_2\frac{\epsilon_2}{p_1p_3}+\ii\alpha(a)-\frac{\omega_4}{p}\frac{\alpha(\mathfrak{m}_1)}{|\boldsymbol{p}|}+\frac{\R_3}{p_1p_2}\alpha(\mathfrak{m}_2)\right)}{\prod\limits_{(m_1,m_2)\in\mathcal{B}^{+-}_{\widehat{B}}(\mathfrak{m}_1,\mathfrak{m}_2)}\left(m_1\frac{\epsilon_1}{p_3}+m_2\frac{\epsilon_2}{p_1p_3}+\ii\alpha(a)-\frac{\omega_4}{p}\frac{\alpha(\mathfrak{m}_1)}{|\boldsymbol{p}|}+\frac{\R_3}{p_1p_2}\alpha(\mathfrak{m}_2)\right)}\\
    &\frac{\prod\limits_{(m_1,m_2)\in\mathcal{B}^{--}_{\widehat{B}}(\mathfrak{m}_1,\mathfrak{m}_2)}\left(m_1\frac{\epsilon_1}{p_3}+m_2\frac{\epsilon_2}{p_1p_3}+\ii\alpha(a)-\frac{\omega_4}{p}\frac{\alpha(\mathfrak{m}_1)}{|\boldsymbol{p}|}+\frac{\R_3}{p_1p_2}\alpha(\mathfrak{m}_2)\right)}{\prod\limits_{(m_1,m_2)\in\mathcal{B}^{-+}_{\widehat{B}}(\mathfrak{m}_1,\mathfrak{m}_2)}\left(m_1\frac{\epsilon_1}{p_3}+m_2\frac{\epsilon_2}{p_1p_3}+\ii\alpha(a)-\frac{\omega_4}{p}\frac{\alpha(\mathfrak{m}_1)}{|\boldsymbol{p}|}+\frac{\R_3}{p_1p_2}\alpha(\mathfrak{m}_2)\right)},
\end{split}\end{equation}
where
\begin{equation}\setlength{\jot}{10pt}\begin{split}\label{eq.Bpmpm}
    &\mathcal{B}^{++}_{\widehat{B}}=\text{proj}_{12}\left\{\vec{m}\in\mathcal{C}^{++}_{\widehat{Y}}\,|\,m_3=t_{\boldsymbol{p}}\right\},\qquad\quad\mathcal{B}^{+-}_{\widehat{B}}=\text{proj}_{12}\left\{\vec{m}\in\mathcal{C}^{+-}_{\widehat{Y}}\,|\,m_3=t_{\boldsymbol{p}}\right\},\\
    &\mathcal{B}^{-+}_{\widehat{B}}=\text{proj}_{12}\left\{\vec{m}\in\mathcal{C}^{-+}_{\widehat{Y}}\,|\,m_3=t_{\boldsymbol{p}}\right\},\qquad\quad\mathcal{B}^{--}_{\widehat{B}}=\text{proj}_{12}\left\{\vec{m}\in\mathcal{C}^{--}_{\widehat{Y}}\,|\,m_3=t_{\boldsymbol{p}}\right\}.
\end{split}\end{equation}
Expressions for the one-loop determinant of hypermultiplets at a generic flux sector $(\mathfrak{m}_1,\mathfrak{m}_2)$ follow from those presented above.

\subsection{Examples}
Focusing on vector multiplets, we now present two explicit examples: $Y^{2,1}/S^1$ and $Y^{3,1}/S^1$. We present the lattices of charges $(m_1,m_2)$ contributing to the one-loop determinant at a generic flux sector $(\mathfrak{m}_1,\mathfrak{m}_2)$ \eqref{eq.Btopex} and \eqref{eq.Bpmpm}. The lattices contributing to theories with twist defects on $\widehat{B}_{\boldsymbol{p}}$ are similar. To draw them it is enough to consider the rescaling of $\alpha(\mathfrak{m}_i)$ depending on $(p_1,\dots,p_4)$. However, notice that in the one-loop determinant \eqref{eq.1loop4.ex.fin} also the equivariant parameters $\epsilon_1,\epsilon_2$ are affected by the parameters $(p_1,\dots,p_4)$.

We start considering the case with $\alpha(\mathfrak{m}_1)=0$: the lattice in \eqref{eq.Btopex} contributing to the partition functions \eqref{eq.1loop4.ex} are shown in \autoref{fig.1} for various values of $\mathfrak{m}_2$.
\begin{figure}[h!]
\centering
\tdplotsetmaincoords{0}{0}
\begin{tikzpicture}[scale=0.55,tdplot_main_coords]
    \filldraw[ draw=green,fill=green!20,opacity=0.5]       
    (0,0)
    -- (2,-2)
    -- (6,-4)
    -- (6,5)
    -- (0,5)
    -- cycle;
   	\filldraw[ draw=blue,fill=blue!20,opacity=0.5]       
    (0,0)
    -- (6,-3)
    -- (6,5)
    -- (0,5)
    -- cycle;
    \filldraw[ draw=red,fill=red!20,opacity=0.5]       
    (0,2)
    -- (3,-1)
    -- (6,-2.5)
	-- (6,5)
    -- (0,5)
	-- cycle;
    \draw[step=1.0,gray!60] (-.5,-5.5) grid (6.5,5.5);
	\draw[thick,-stealth] (-0.5,0) -- (6.5,0) node[anchor=south]{$m_1$};
    \draw[thick,-stealth] (0,-5.5) -- (0,5.5) node[anchor=east]{$m_2$};
        \draw[thick,green] 
    (0,5)
    -- (0,0)
    -- (2,-2)
	-- (6,-4);
    \draw[thick,blue]
    (0,5)
    -- (0,0)
    -- (6,-3);
    \draw[thick,red] 
    (0,5)
    -- (0,2)
    -- (3,-1)
    -- (6,-2.5);
\end{tikzpicture}
\hspace{6em}
\begin{tikzpicture}[scale=0.55,tdplot_main_coords]
    \filldraw[ draw=green,fill=green!20,opacity=0.5]       
    (0,0)
    -- (4,-4)
    -- (6,-5)
    -- (6,5)
    -- (0,5)
    -- cycle;
   	\filldraw[ draw=blue,fill=blue!20,opacity=0.5]       
    (0,0)
    -- (6,-3)
    -- (6,5)
    -- (0,5)
    -- cycle;
    \filldraw[ draw=red,fill=red!20,opacity=0.5]       
    (0,3)
    -- (4,-1)
    -- (6,-2)
	-- (6,5)
    -- (0,5)
	-- cycle;
    \draw[step=1.0,gray!60] (-.5,-5.5) grid (6.5,5.5);
	\draw[thick,-stealth] (-0.5,0) -- (6.5,0) node[anchor=south]{$m_1$};
    \draw[thick,-stealth] (0,-5.5) -- (0,5.5) node[anchor=east]{$m_2$};
    \draw[thick,green] 
    (0,5)
    -- (0,0)
    -- (4,-4)
	-- (6,-5);
    \draw[thick,blue]
    (0,5)
    -- (0,0)
    -- (6,-3);
    \draw[thick,red] 
    (0,5)
    -- (0,3)
    -- (4,-1)
    -- (6,-2);
\end{tikzpicture}
\caption{Lattices $\mathcal{B}_B$ for $\alpha(\mathfrak{m}_1)=0$. Left side: $Y^{2,1}$ for $\alpha(\mathfrak{m}_2)=-1$ (red), $\alpha(\mathfrak{m}_2)=0$ (blue) and $\alpha(\mathfrak{m}_2)=2$ (green). Right side: $Y^{3,1}$ for $\alpha(\mathfrak{m}_2)=-1$ (red), $\alpha(\mathfrak{m}_2)=0$ (blue) and $\alpha(\mathfrak{m}_2)=2$ (green).}
\label{fig.1}
\end{figure} 
As expected for Pestun-like theories, the latices are non-compact. For $p=2,\,q=1$ these plots reproduce those appearing in \cite{Lundin:2023tzw} where the dimensional reduction only of the perturbative partition function on $Y^{p,q}$ has been considered.

We now consider $\alpha(\mathfrak{m}_2)=0$ with generic $\alpha(\mathfrak{m}_1)$ and we plot in \autoref{fig.2} the lattices $\mathcal{B}_B^{++}$, $\mathcal{B}_B^{+-}$, $\mathcal{B}_B^{--}$ and $\mathcal{B}_B^{-+}$\eqref{eq.Bpmpm} contributing to \eqref{eq.1loop4.ex.fin}.
\begin{figure}[h!]
\centering
\tdplotsetmaincoords{0}{0}
\begin{tikzpicture}[scale=0.55,tdplot_main_coords]
   	\filldraw[ draw=red,fill=red!20,opacity=0.5]       
    (0,2)
    -- (5,-0.5)
    -- (5,5)
    -- (0,5)
    -- cycle;
    \filldraw[ draw=blue,fill=blue!20,opacity=0.5]       
    (0,0)
    -- (0,2)
    -- (-4,4)
    -- cycle;
    \filldraw[ draw=green,fill=green!20,opacity=0.5]       
    (0,0)
    -- (0,-5)
    -- (-5,-5)
    -- (-5,4.5)
    -- (-4,4)
    -- cycle;
    \draw[step=1.0,gray!60] (-5.5,-5.5) grid (5.5,5.5);
	\draw[thick,-stealth] (-5.5,0) -- (5.5,0) node[anchor=south]{$m_1$};
    \draw[thick,-stealth] (0,-5.5) -- (0,5.5) node[anchor=east]{$m_2$};
    \draw[thick,red]
    (0,5)
    -- (0,2)
    -- (5,-0.5);
    \draw[thick,green]
    (0,-5)
    -- (0,0)
    -- (-4,4)
    -- (-5,4.5);
    \draw[thick,blue]
    (0,0)
    -- (0,2)
    -- (-4,4)
    -- cycle;
\end{tikzpicture}
\hspace{2em}
\begin{tikzpicture}[scale=0.55,tdplot_main_coords]
    \filldraw[ draw=red,fill=red!20,opacity=0.5]       
    (0,0)
    -- (0,5)
    -- (5,5)
    -- (5,-4.5)
    -- (4,-4)
    -- cycle;
    \filldraw[ draw=brown,fill=yellow!20,opacity=0.5]       
    (0,0)
    -- (0,-2)
    -- (4,-4)
    -- cycle;
    \filldraw[ draw=green,fill=green!20,opacity=0.5]       
    (0,-2)
    -- (-5,0.5)
    -- (-5,-5)
    -- (0,-5)
    -- cycle;
    \draw[step=1.0,gray!60] (-5.5,-5.5) grid (5.5,5.5);
	\draw[thick,-stealth] (-5.5,0) -- (5.5,0) node[anchor=south]{$m_1$};
    \draw[thick,-stealth] (0,-5.5) -- (0,5.5) node[anchor=east]{$m_2$};
    \draw[thick,red]
    (0,5)
    -- (0,0)
    -- (4,-4)
    -- (5,-4.5);
    \draw[thick,yellow]
    (0,0)
    -- (0,-2)
    -- (4,-4)
    -- cycle;
    \draw[thick,green]
    (0,-5)
    -- (0,-2)
    -- (-5,0.5);
\end{tikzpicture}
\caption{Lattices $\mathcal{B}_Y^{++},\,\mathcal{B}_Y^{--}$ and $\mathcal{B}_Y^{+-},\,\mathcal{B}_Y^{-+}$, for $\alpha(\mathfrak{m}_2)=0$ on $Y^{2,1}$. Left side: $\mathcal{B}_Y^{++}$ (red), $\mathcal{B}_Y^{+-}$ (blue) and $\mathcal{B}_Y^{--}$ (green) for $\alpha(\mathfrak{m_1})=8$. Right side: $\mathcal{B}_Y^{++}$ (red), $\mathcal{B}_Y^{-+}$ (yellow) and $\mathcal{B}_Y^{--}$ (green) $\alpha(\mathfrak{m_1})=-8$.}
\label{fig.2}
\end{figure}
Recall the contributions in  $\mathcal{B}_Y^{++},\,\mathcal{B}_Y^{--}$ appear at the numerator in the one-loop determinant while $\mathcal{B}_Y^{+-},\,\mathcal{B}_Y^{+-}$ appear at the denominator. Let us stress that such contributions at the denominator arise only for $\alpha(\mathfrak{m}_1)\neq 0$ and thus, they did not appear in earlier works \cite{Lundin:2023tzw}.

\section{Factorized Expressions and Instantons}\label{sec.5}
We show how the one-loop determinant at different flux sectors, both in 5d and 4d, can be computed gluing contributions from different fixed fibers in 5d and different fixed points in 4d. This factorization property was shown to hold for the perturbative partition function in 5d \cite{Qiu:2014oqa} and in 4d \cite{Festuccia:2019akm,Mauch:2021fgc}, and for the one-loop determinant around $\mathfrak{m}_2$ on quasi-toric four-manifolds \cite{Lundin:2023tzw} and for the 4d theory obtained dimensionally reducing from a branched cover of $S^5$ \cite{Mauch:2024uyt}. In this section we only treat vector multiplets without insertion of twist defects. However, the same steps continue to hold also for hypermultiplets, and on $\widehat{Y}^{p,q}_{\boldsymbol{p}}$ and $\widehat{B}_{\boldsymbol{p}}$; the required modifications are along the lines as those appearing in \cite{Mauch:2024uyt}.

\paragraph{Five Dimensions.}
The factorized expression for the one-loop determinant on $Y^{p,q}$ \eqref{eq.1loop.red.n} depends on a choice of imaginary part of\footnote{Each individual contribution is affected by this choice. However, after gluing all contributions, the partition function is independent.} $\omega_i$. The following rewriting holds
\begin{equation}\label{eq.fact1}
    Z_{Y^{p,q}}^\text{vm}=\prod_{\alpha\in\Delta}\prod_{i=1}^4 \prod_{t}\Upsilon_i( \ii\alpha(a)+\beta_{1i}^{-1}\alpha(\mathfrak{m}_1) +\beta_{2i}^{-1}t|\epsilon^i_1,\epsilon^i_2)^{s_i},
\end{equation}
where we defined the $\Upsilon$-function \cite{Hama:2012bg,Festuccia:2018rew,Lundin:2023tzw}
\begin{equation}
    \Upsilon_i(z|\epsilon_1,\epsilon_2)=\prod_{(j,k)\in\mathcal{D}_i}(\epsilon_1 j+\epsilon_2 k+z)\prod_{(j,k)\in\mathcal{D}_i^\prime}(\epsilon_1 j+\epsilon_2 k+\bar{z}).
\end{equation}
The product over four contributions in \eqref{eq.fact1} is from neighbourhoods $\mathbb{C}^2_{\epsilon_1^i,\epsilon_2^i}\times S^1$ around each fixed fiber of $Y^{p,q}$. The local equivariant parameters for the $T^3$-action are $\epsilon_1^i,\,\epsilon
_2^1,\,\beta_{2i}^{-1}$. These, together with the shift in front of $\alpha(\mathfrak{m}_2)$ and the parameters $s_i$ can be read in \autoref{tab.1}. 
\begin{table}[h]
\centering
\begin{tabular}{ c || c | c | c | c }
    $i$ & 1 & 2 & 3 & 4\\ \hline
    & & & & \\[\dimexpr-\normalbaselineskip+3pt]
    $\epsilon_1^i$ & $\epsilon_1-\epsilon_2$ & $2\epsilon_1-\epsilon_1$ & $\epsilon_2-\epsilon_1$ & $\epsilon_2$\\ 
    $\epsilon_2^i$ &  $\epsilon_2$ & $\epsilon_2-\epsilon_1$ & $2\epsilon_1-\epsilon_2 $ & $\epsilon_1-\epsilon_2$\\  
    $\beta_{1i}^{-1}$ & $\frac{\omega_4}{p}$ &  $\frac{\omega_4}{p}$ & $\frac{\omega_4}{p}-\frac{\epsilon_2}{p+q}$ & $\frac{\omega_4}{p}-\frac{\epsilon_2}{p+q}$\\
    $\beta_{2i}^{-1}$ & $\R_3$ &  $(p-q)(\epsilon_1-\epsilon_2)+\R_3 $ & $ (p+q)\epsilon_1-q\epsilon_2-\R_3 $ & $p\epsilon_2-\R_3 $\\
    $s_i$ & $ -1 $ & $ -1 $ & $ 1 $ & $1$
\end{tabular}
\caption{Local equivariance parameters in 5d and in 4d}
\label{tab.1}
\end{table}
The lattices $\mathcal{D}_i,\mathcal{D}_i'$, defined in \cite{Lundin:2023tzw}, and the parameters $s_i$, are determined by different choices of regularization at the fixed fibers. Once the local equivariant parameters are known also on $L^{a,b,c}$, the factorization takes the same form as in \eqref{eq.fact1}.

As on quasi-toric four-manifolds \cite{Nekrasov:2003vi,Festuccia:2018rew,Lundin:2023tzw}, also in 5d the dependence on $\mathfrak{m}_1$ is via a shift of the Coulomb branch parameter. This allows us to conjecture the contact instanton contributions to the full partition function on $Y^{p,q}$ \eqref{eq.Zfull.M} to be the following
\begin{equation}
    Z_{Y^{p,q}}^\text{inst}=\prod_{i=1}^4 Z^{Nek}_{\mathbb{C}^2\times S^1}(\ii a+\beta_{1i}^{-1}\mathfrak{m}_1|\beta_{2i}^{-1},\epsilon^i_1,\epsilon^i_2,q),
\end{equation}
where $q=\exp(2\pi\ii\tau)$ is the instanton counting parameter.

\paragraph{Four Dimensions.}
Once the factorization of the one-loop determinant on $Y^{p,q}$ is determined, the one-loop contributions on $B$ follow simply by
restricting $t$
\begin{equation}\label{eq.fact2}
    Z^\text{vm,ex}_{B}=\prod_{\alpha\in\Delta}\prod_{i=1}^4 \Upsilon_i( \ii\alpha(a)+\beta_{1i}^{-1}\alpha(\mathfrak{m}_1)+\beta_{2i}^{-1}\alpha(\mathfrak{m}_2)|\epsilon^i_1,\epsilon^i_2)^{s_i}.
\end{equation}
The local equivariant parameters appear in \autoref{tab.1}. The flux dependence enters as a shift of the Coulomb branch parameter for both $\mathfrak{m}_1$ and $\mathfrak{m}_2$. Therefore, we can write down the instanton contributions to the full partition function \eqref{eq.Zfull.B} on $B$
\begin{equation}
    Z_{B}^\text{inst,ex}=\prod_{i=1,4} Z^{Nek}_{\mathbb{C}^2}(\ii a+\beta_{1i}^{-1}\mathfrak{m}_1+\beta_{2i}^{-1}\mathfrak{m}_2|\epsilon^i_1,\epsilon^i_2,q)\prod_{i=2,3} Z^{Nek}_{\mathbb{C}^2}(\ii a+\beta_{1i}^{-1}\mathfrak{m}_1+\beta_{2i}^{-1}\mathfrak{m}_2|\epsilon^i_1,\epsilon^i_2,\bar{q}).
\end{equation}
Note that the distribution of (anti-)instantons at different fixed points reflects the fact that we are considering an exotic, or Pestun-like, theory.

\section{Discussion}\label{sec.6}
In this work we computed partition functions of $\mathcal{N}=1$ theories on five-dimensional toric Sasakian manifolds $Y^{p,q}$ and $L^{a,b,c}$, including previously neglected contributions from gauge configurations with flux. We extended these results to theories with twist defects on $\widehat{Y}^{p,q}_{\boldsymbol{p}}$ and $\widehat{L}^{a,b,c}_{\boldsymbol{p}}$. According to the proposal in \cite{Mauch:2025irx}, for vector multiplets and hypermultiplets in the adjoint representation, the partition function on these spaces is reproduces that on orbifolds $Y^{p,q}_{\boldsymbol{p}}$ and $L^{a,b,c}_{\boldsymbol{p}}$.

We further employed these results to compute the full partition function for an $\mathcal{N}=2$ SQFT on a manifold $B$ whose topology is that of the product of two spheres, and on the corresponding orbifold. In particular, with the same caveat as above, our results provide the partition function on a class of orbifolds whose topology is that of the product of two spindles.

We computed all these results exploiting a sequence of non-trivial $S^1$-fibrations, that is $Y^{p,q}=(S^3\times S^3)/S^1_Y$ (similar for $L^{a,b,c}$) and $B=Y^{p,q}/S^1$. Thus, we have found that the equivariant index of a certain complex on $S^3\times S^3$ completely determines the one-loop determinant around both fluxes on $B$. The same holds on orbifolds, starting from the equivariant index on $\widehat{S}^3_{\boldsymbol{p}_A}\times \widehat{S}^3_{\boldsymbol{p}_B}$.

\paragraph{Future Directions.}
\begin{itemize}
    \item It would be interesting to compute all flux contributions on a generic toric Sasakian manifolds. Once such a result is know, it would be possible to dimensionally reduce to quasi-toric manifolds, along the lines of \cite{Lundin:2023tzw,Mauch:2024uyt} for a single flux sector. In general, the number of flux sectors is $m-3$ in 5d and $m-2$ in 4d, where $m$ appears in \eqref{eq.Kquotient}.  
    \item Recently, the large $N$ limit of the spindle index \cite{Inglese:2023wky,Inglese:2023tyc} has been shown to reproduce the entropy of certain accelerating black holes \cite{Colombo:2024mts,BenettiGenolini:2024hyd}. A technical difficulty in their computations is that fluxes for the gauge field contain a fractional part, which needs to be carefully treated. The entropy has also been computed for AdS$_6$ supergravity solutions with near horizon geometry given by a space whose topology is AdS$_2\times \mathbb{CP}^1_{\boldsymbol{p}_A}\times \mathbb{CP}^1_{\boldsymbol{p}_B}$ \cite{Faedo:2024upq,Couzens:2025ghx}. This entropy is expected to be reproduced by the large $N$ partition function of \emph{Seiberg theories} \cite{Seiberg:1996bd}, that is $\mathcal{N}=1$ $USp(2N)$ theories with $N_f$ flavours and an antisymmetric matter field, on $\mathbb{CP}^1_{\boldsymbol{p}_A}\times \mathbb{CP}^1_{\boldsymbol{p}_B}\times S^1$. This partition function is obtained by taking as building block \eqref{eq.1loop4.ex.fin} and the corresponding expression for a hypermultiplet\footnote{Let us stress, once more, that even if the partition function in \autoref{sec.4} is computed for matter in a generic representation $R$, the equivalence in \cite{Mauch:2025irx} holds only for the adjoint representation. In order to use these results in other cases one would have to prove that the equivalence holds in general.}, and adding the contributions of modes along the extra $S^1$. For smooth manifolds this has been studied in \cite{Crichigno:2018adf,Hosseini:2018uzp}.
    \item It is known that the partition function of $\mathcal{N}=1$ SQCD with $SU(2)$ gauge group and four flavours on $S^5$, tuning the masses to specific values, reproduces the partition function of SQED with four flavours on $S^3$ \cite{Gaiotto:2012xa,Nieri:2013vba}. In particular, two of the three contributions from the fixed fibers of $S^5$ reduce to vortex partition functions, while the third trivializes. The two remaining contributions pair up to give the partition function on $S^3$ \cite{Pasquetti:2011fj,Beem:2012mb}, corresponding to a facet of the toric diagram of $S^5$. It would be interesting to generalize this result to $Y^{p,q}$ where one expects that the partition function reduces, at specific values of the masses, to the partition function on one of the facets of the corresponding toric diagram, that is a lens space.
\end{itemize}

\paragraph*{Acknowledgments}
        We are grateful to Roman Mauch for stimulating discussions and enjoyable collaboration on related topics. LR acknowledges support from the Shuimu Tsinghua Scholar Program.

\bibliographystyle{utphys}
\bibliography{main}

\end{document}